\documentclass[11pt,oneside,letterpaper]{article}
\usepackage{eurosym}
\usepackage{graphicx,amsmath,amsfonts}
\usepackage{setspace}
\usepackage{amssymb}
\usepackage{dsfont}
\usepackage{amsmath}
\usepackage{amsfonts}
\usepackage{ifthen}
\usepackage{mathtools}
\usepackage{graphicx}
\usepackage{enumerate}
\usepackage{enumitem}
\usepackage{color}
\usepackage{framed}
\usepackage[margin=1in]{geometry}
\usepackage{natbib}
\usepackage[dvipsnames]{xcolor}
\PassOptionsToPackage{hyphens}{url}
\usepackage[colorlinks=true, urlcolor=Mahogany, linkcolor=Mahogany, citecolor=Mahogany]{hyperref}
\usepackage{fp}
\usepackage[capitalize]{cleveref}
\usepackage{amsthm}
\usepackage{caption}
\usepackage{subcaption}
\usepackage{palatino}
\usepackage{cuted,balance}
\usepackage{amssymb}
\usepackage{soul}
\usepackage{graphicx,amsmath,amsfonts}

\DeclareMathOperator*{\argmax}{arg\,max}
\newtheorem{assumption}{Assumption}

\newtheorem{corollary}{Corollary}
\newtheorem{definition}{Definition}
\newtheorem{proposition}{Proposition}
\newtheorem{theorem}{Theorem}
\newtheorem{lemma}{Lemma}

\newtheorem{remark}{Remark}

\allowdisplaybreaks

\let\originaleqref\eqref
\renewcommand{\eqref}{\originaleqref}

\usepackage{bm}
\newcommand{\btheta}{\bm{\theta}}
\newcommand{\bx}{\bm{x}}
\newcommand{\by}{\bm{y}}
\newcommand{\ba}{\bm{a}}
\newcommand{\bb}{\bm{b}}
\newcommand{\be}{\bm{e}}
\newcommand{\bsigma}{\bm{\sigma}}
\newcommand{\bgamma}{\bm{\gamma}}
\newcommand{\bp}{\bm{p}}
\newcommand{\bc}{\bm{c}}
\newcommand{\bts}{\bm{\tilde{s}}}
\newcommand{\tsnn}{\tilde{\Sigma}_{(0,0)}}
\newcommand{\tsyn}{\tilde{\Sigma}_{(1,0)}}
\newcommand{\tsny}{\tilde{\Sigma}_{(0,1)}}
\newcommand{\tsyy}{\tilde{\Sigma}_{(1,1)}}

\begin{document}

\title{On Three-Layer Data Markets}

\author{Alireza Fallah \thanks{Department of Computer Sciences and Ken Kennedy Institute, Rice University} \and 
Michael I. Jordan\thanks{INRIA Paris and University of California, Berkeley} \and Ali Makhdoumi \thanks{%
Fuqua School of Business, Duke University} \and Azarakhsh Malekian \thanks{%
Rotman School of Management, University of Toronto.}}

\date{November 2025}
\maketitle

\begin{abstract}
We study a three-layer data market comprising users (data owners), platforms, and a data buyer. Each user benefits from platform services in exchange for data, incurring privacy loss when their data, albeit noisily, is shared with the buyer. The user chooses platforms to share data with, while platforms decide on data noise levels and pricing before selling to the buyer. The buyer selects platforms to purchase data from. We model these interactions via a multi-stage game, focusing on the subgame Nash equilibrium. We find that when the buyer places a high value on user data (and platforms can command high prices), all platforms offer services to the user who joins and shares data with every platform. Conversely, when the buyer's valuation of user data is low, only large platforms with low service costs can afford to serve users. In this scenario, users exclusively join and share data with these low-cost platforms. Interestingly, increased competition benefits the buyer, not the user: as the number of platforms increases, the user utility does not necessarily improve while the buyer utility improves. However,  increasing the competition improves the overall utilitarian welfare. Building on our analysis, we then study regulations to improve the user utility. We discover that banning data sharing maximizes user utility only when all platforms are low-cost. In mixed markets of high- and low-cost platforms, users prefer a minimum noise mandate over a sharing ban. Imposing this mandate on high-cost platforms and banning data sharing for low-cost ones further enhances user utility.

\end{abstract}




\section{Introduction}
In the digital era, online platforms have become integral to our daily lives, offering a plethora of services that range from social networking to personalized shopping experiences. While these platforms provide convenience and connectivity, they also engage in extensive data collection practices. Users' interactions with these platforms generate vast amounts of data, encompassing personal preferences, behaviors, and other sensitive information. This data, a valuable commodity, is often shared with third-party buyers.

With the advent of new functionalities by information technology companies such as Apple or Google that reduce or eliminate tracking via third-party cookies, there has been an increased reliance on \textit{first-party} data, which is information gathered by a company from its own user base. For example, as elaborated by \cite{cross2023new}, major retailers such as BestBuy and Walmart capitalize on this data through loyalty programs, customer purchase records, and subscriptions. They do this by selling the data to advertisers who aim to place targeted ads on the retailers' websites or across various online platforms. In a similar vein, \cite{Cross_2023} notes that Mastercard, a leading payment technology firm, trades cardholder transaction information via third-party online data marketplaces and its internal ``Data and Services" division. This practice grants numerous entities access to extensive consumer data and insights. For instance, their Intelligent Targeting service allows businesses to leverage ``Mastercard 360$^{\circ}$ data insights" to create and execute advertising campaigns that specifically target potential ``high-value" customers.

In this complex digital ecosystem, users face the dilemma of benefiting from the services provided by these platforms while potentially compromising their privacy. Platforms, on the other hand, must balance the profitability derived from data sharing with the need to maintain user trust and satisfaction. This interplay between users' privacy concerns and the economic incentives of platforms raises important questions about the sustainability of current data market practices. Notable regulatory actions have been taken to improve user experiences. For instance, the General Data Protection Regulation (GDPR) and the California Consumer Privacy Act (CCPA) have been introduced to protect user privacy and to limit the amount of data shared with the platforms. Although, at first sight, this helps users, there have been several studies pointing to its detrimental impact both on platforms and users. This is because these regulations have led to a more concentrated market structure by limiting competition in data markets. This results from the regulation's restrictions on data sharing, which may slow down data-based innovations and services. For instance, studies have shown that GDPR has led to a significant reduction in the number of available apps in the market, with new app entries falling by half following the regulation's implementation \citep[see, e.g.,][]{aridor2021effect,janssen2022gdpr}. This decline in market entries indicates a potential loss of innovation as new and smaller players find it increasingly difficult to compete under these regulations. This effect is particularly detrimental to small companies that rely on these data synergies for innovation and market growth \citep{gal2020competitive}. {On a similar note, \cite{demirer2024data} also show that GDPR, on average, represents a 20\% increase in the cost of data, with the ``distortionary effects of the GDPR are highest for the smallest firms."}

In this paper, we aim to understand the user and platform strategies that arise when considering the collection and usage of user data. We also explore regulatory frameworks that could be implemented to enhance user utility while maintaining the benefits provided by online platforms.

\subsection{Our model and main results}
We adopt an analytical-modeling-based approach and develop a framework to examine the choices of both users and platforms, taking into account users' privacy concerns. We also explore regulations and policies that could alleviate these concerns. In our model, multiple platforms interact with a user who owns data and seeks the services of these platforms, as well as with a third-party buyer interested in purchasing the user's data from the platforms. Throughout our study, we consider MasterCard as an example of such a platform. The user represents an individual in a specific population segment with particular preferences, and the buyer is conceptualized as a third-party advertiser aiming to understand and leverage these user preferences for its own benefit.

The utility of each platform is determined by three factors: the gains obtained from using user data to provide services, potential payments received from buyers purchasing this data, and the costs incurred in providing services to the user. Let us highlight the cost of services, as it plays an important role in our analysis. It is particularly important to distinguish between \emph{high-cost} and \emph{low-cost} platforms. The low-cost platforms are those that do not necessarily need to sell data to stay in the market, as their gains from the service already cover their costs of operation. In our leading examples, low-cost platforms are exemplified by payment-technology companies or big retailers, where either their operational costs per user are low, or their main source of revenue is not derived from selling data. Conversely, for high-cost platforms, the gains from service provision are not enough to offset their operational costs. Hence, they require a minimum level of data selling to be incentivized to enter the market (i.e., to provide services to the user). Examples of these platforms typically include smaller firms or companies that heavily rely on data collection for advertising to ensure their continued operation.

We model the interaction between the players as a multi-stage game in which, in the first stage, the platforms decide whether they want to provide services to the user and the privacy level they want to deliver to the data buyer (so that the user is incentivized to share). We model the privacy level that the platform provides as a \emph{noise level} added to the user data. Adding a suitable level of noise is the common practice for achieving privacy, primarily motivated by the differential privacy framework of \cite{dwork2014algorithmic}. This approach is widely used both in the machine learning literature and in real-world applications, ranging from census data \citep[see, e.g.,][]{abowd2018us} to technology companies such as Google and Apple \citep[see, e.g.,][]{erlingsson2014rappor, apple}.


In the second stage of the game, the user decides which platform to join. The user's utility is the service quality of the platform, proportional to the amount of user information revealed to the platform minus the user's \textit{privacy loss}. The privacy loss refers to the amount of information all platforms reveal to the data buyer. In the third and fourth stages, the platforms determine the price to charge the data buyer, and the buyer decides from which platforms to purchase. The data buyer's utility is the user's leaked information from all platforms minus the payments made to acquire this information. Again, the data buyer can be viewed as a third-party advertiser in our example mentioned above who pays the platforms to obtain user data. They then use this data for potential gain, with their gain being proportional to the accuracy of the user data obtained.

\paragraph*{Equilibrium characterization and insights:} We adopt the subgame perfect equilibrium as our equilibrium concept and use backward induction to characterize it. Finding the equilibrium of this game is complex mainly because the platforms' choice of noise level not only directly impacts the user's utility but also changes the data buyer's decision, which indirectly impacts the user's utility. Also, the competition among platforms further complicates the dynamic of the game. For instance, a platform may still be incentivized to sell data at a noise level higher than needed for the user to be willing to share their data. At first glance, this may seem unreasonable as it reduces the quality of the data that they can sell, and hence, the data buyer would pay less for it. However, increasing the privacy could persuade the user to forgo other platforms that offer less stringent privacy guarantees and only use this platform's service. As a result, this can give a monopoly to the platform in terms of access to the user's data, and hence, the data buyer would end up paying more to them.

We start our analysis of equilibrium by focusing on the case of two platforms. This setting allows us to infer insights about equilibrium properties that are applicable to more general cases. Here, we identify up to three primary regimes, depending on the data buyer's valuation of user data acquisition, denoted by $\beta$ in our model. When $\beta$ falls below a certain threshold, neither of the two platforms enters the market due to insufficient payment to cover service costs. This regime ceases to exist when at least one platform is low-cost, meaning it can cover service costs solely through service provision without needing to sell data. Conversely, when $\beta$ is sufficiently high, both platforms will participate in the market. Interestingly, in this case, and when the user is privacy-conscious and requires a minimum non-zero level of noise for data sharing, the user utility is equal to zero at equilibrium. However, as the data buyer compensates each platform at the marginal value of the data, their utility remains positive. Finally, there exists an intermediate regime of $\beta$, where only the lower-cost platform engages in the market. Here, both the user and the data buyer utilities are zero at equilibrium, indicating that having both platforms in the market to compete benefits the data buyers but not the user.

We demonstrate that these findings extend to the general scenario with $K>2$ platforms. Specifically, we establish that when $\beta$ exceeds a certain threshold, signifying the data buyer's willingness to pay more for the user data, all platforms will enter the market and provide services to the user. Conversely, only low-cost platforms will participate when $\beta$ is below this threshold, while high-cost platforms will abstain. Although competition does not favor the user, it results in positive utility for the data buyer. Moreover, the overall utilitarian welfare increases proportionally with the number of participating platforms.

\paragraph*{Regulation and policy design:} Building on our equilibrium characterization, we then study regulations that aim to improve user utility. Here, our analysis provides three main insights. Firstly, we will consider a ban on platforms sharing user data with data buyers. This regulation only maximizes the user utility if all the platforms are low-cost. Specifically, under this regulation, users obtain the best of both worlds: all platforms (that have a low cost for providing services) enter the market, and the user benefits from the services of all of them while incurring zero privacy loss. This observation suggests that if we only had large firms whose costs of providing service per user are small, then the draconian regulation that involves banning data sharing would have been optimal. However, this is no longer the case if the market includes both high- and low-cost platforms.

In this case, our second insight reveals that imposing a uniform minimum privacy mandate improves user utility, particularly when the value of the user data to the data buyer is significant enough for platforms to charge a high price for it. This means that all platforms need to perturb the user data by adding to it a noise with a high enough noise level. Here, our analysis identifies specific conditions in which GDPR-type policies, which limit user data usage, can benefit users. Finally, our third insight suggests that, despite potential implementation challenges, a non-uniform minimum privacy mandate—banning data sharing for low-cost platforms while imposing a minimum privacy standard for high-cost platforms—further enhances user utility. This regulation prohibits large, low-cost platforms from sharing user data while permitting smaller, high-cost platforms to do so, albeit in a limited manner through privacy mechanisms. This approach aligns with our earlier discussion that GDPR-type regulations disproportionately harm small businesses. Therefore, a non-uniform regulation favoring small businesses is preferred not only by these businesses but also by the users.

\subsection{Related literature}

Our paper broadly relates to the emerging literature on data markets and online platform behavior. Earlier works on this topic include \cite{acquisti2016economics}, \cite{posner2018radical}, 
\cite{ali2019voluntary}, \cite{jones2020nonrivalry}, and \cite{dosis2022}, all of whom study the consequences of protecting and disclosing personal information about individuals. Additionally, \cite{acemoglu2019too},  \cite{bergemann2020economics},  \cite{ichihashi2020online}, and \cite{fainmesser2022digital} study the privacy consequences of data externality (whereby a
user's data reveals information about others). \cite{acemoglu2023good} study user-optimal privacy-preserving mechanisms that platforms can adopt to balance privacy and learning and \cite{acemoglu2023model}  develop an experimentation game to study the implications of the platform's information advantage in product offering. Finally, \cite{ichihashi2021competing} studies data intermediaries, and \cite{ichihashi2020dynamic} studies consumer privacy choices while online platforms dynamically adjust their privacy guarantees to incentivize more consumer data sharing.

More closely related to ours are papers that study data sharing among users, platforms, and third-party data buyers and, in particular, the impacts of banning such data sharing. In this regard, \cite{bimpikis2021data} and \cite{Argenziano2023} develop a two-round model of the interaction between a user and two competing platforms that can collect data on user preferences (and potentially use it for price discrimination). \cite{bimpikis2021data} show that banning data sharing can hurt the user when the platforms offer complementary products, and \cite{Argenziano2023} show that banning data sharing can hurt the user if the benefits of knowing user preferences by the seller and personalized services are limited.  The above papers consider the interactions between a user and a platform that itself uses the user data for price discrimination and, therefore, can potentially hurt the user. In contrast, we study a different problem that relates to how platforms sell user data to other third-party data buyers. Therefore, in addition to the differences in modeling choices and analysis, we depart from the above papers by considering a three-layer data market that includes the interaction between the platforms and the data buyer and focusing on deriving insights on effective regulations. Furthermore, \cite{fang2021data} and \cite{bergemann2024data} develop a model to analyze the impact of competition on how data-rich platforms can extract value from sellers through advertising.

Also related to our work is \cite{ravichandran2019effect}, who empirically study the effect of the data-sharing ban on platforms' revenues, and \cite{madsen2023insider}, who study the effects of a complete ban on innovation (we refer to \citet{pino2022microeconomics} for a comprehensive survey on the microeconomics of data). We depart from these papers by focusing on the impact of regulations (and in particular, data-sharing ban) on users' utility. Our paper also relates to the literature that studies various forms of monetizing user data by platforms, such as selling cookies \citep{bergemann2015selling}, efficient pricing for large data sets \citep{abolhassani2017market}, dynamic sales \citep{immorlica2021buying}, \citep{drakopoulos2023providing}, monetization while ensuring the data is replicable \citep{falconer2023replicationrobust}, and the design and price of information where a data buyer faces a decision problem under uncertainty and can
augment their initial private information with supplemental data from a data seller \citep{bergemann2018design}.

More broadly, our paper relates to the growing literature on data acquisition and the design of mechanisms for incentivizing data sharing \citep{ghosh2011selling, ligett2012take, nissim2014redrawing, cummings2015accuracy, chen2018optimal, chen2019prior, fallah2023optimal, cummings2022optimal, fallah2022bridging, karimireddy2022mechanisms}. We depart from this line of work as we consider a three-layer market where the privacy cost occurs due to selling data to a third-party buyer.

{Finally, our paper relates to the supply chain models with competition in the intermediate layer. In these models, upstream and downstream firms interact through contractual arrangements, and the middle layer competes either in prices or quantities (see, e.g., Section 5 of \citep{cachon2003supply}). Related work by \cite{lariviere1999supply}, \cite{corbett2004designing}, and \cite{bernstein2005decentralized} explores how such competition shapes equilibrium prices, efficiency, and welfare. In these settings, competition among intermediaries tends to benefit the downstream buyer and can reduce the intermediaries' profits---an observation that parallels our finding that platform competition benefits the data buyer but not the users.
Despite this structural analogy, our framework departs from the supply chain literature in several key ways. First, the externality in our model arises from \emph{information leakage} and privacy costs rather than inventory or production risk, leading to a fundamentally different welfare trade-off. Second, while classical models feature linear or convex demand and cost functions, the data buyer’s valuation in our model is \emph{submodular} across platforms, capturing diminishing marginal returns to heterogeneous data sources. Finally, instead of price or quantity coordination, the platforms strategically choose their \emph{information quality} (noise levels) and prices, mediating user participation and buyer demand. 

}

The rest of the paper proceeds as follows. In Section \ref{sec:Model}, we describe our model and the interactions among the platforms, the data buyer, and the user. We also present some properties of the revealed information measure in our setting, such as its monotonicity and submodularity. In Section \ref{Sec:Preliminary}, we introduce our equilibrium concept and provide preliminary characterizations of it. In Section \ref{sec:Equilibrium:K=2}, we focus on a setting with two platforms and provide a full characterization of the equilibrium. The focus on two platforms allows us to describe some of the nuances of the interactions in our model. In Section \ref{sec:equilibrium:K>2}, we establish that our main characterizations continue to hold in the general setting with any number of platforms. We also provide several insights and comparative statics, leading to our discussion of the regulations in Section \ref{sec:regulations}. Finally, Section \ref{sec:conclusion} concludes while the appendix contains all the omitted proofs from the text.

\section{Model}\label{sec:Model}

We consider $K \in \mathbb{N}$ \emph{platform}s that are interacting with a \emph{user} and a \emph{data buyer}. The user benefits from using these platforms' services. In return, the platforms acquire data about the user's preferences and behavior. The platforms may subsequently sell this data to a third party, such as an advertiser, which we refer to as a data buyer (or, interchangeably, a buyer). The data buyer is interested in learning about the user's preferences, which can adversely impact the user's utility. This negative impact can be because of price discrimination, unfairly targeted advertising, manipulation, or intrinsic privacy losses. In the example of MasterCard described in the introduction, MasterCard is one of the platforms, and the data buyer is any other firm that purchases user data from MasterCard. 

Formally, we represent the user's data by $ \btheta \in \mathbb{R}^d $, which follows a Gaussian distribution with zero mean and identity variance; that is, $ \btheta \sim \mathcal{N}(0, \mathbb{I}_d) $. This vector $ \btheta $ can be interpreted as the user's feature vector.
Each platform $ i \in [K] \triangleq \{1, \dots, K\} $ offers a service whose quality depends on how accurately the platform can learn the match $\bx_i^\top \btheta$ between the user's feature vector and the platform's characteristics.
Here, $ \bx_i \in \mathbb{R}^d $ is a known unit-norm vector representing the characteristics of platform $ i $. 


The user decides whether to share their data with each of the $K$ platforms in exchange for their service. We let $a_i \in \{0,1\}$ denote whether the user has shared data with platform $i$ and $\ba \in \{0,1\}^K$ \textbf{denote the user strategy profile}. Specifically, if the user opts to share their data with platform $i$, meaning $a_i=1$, then platform $i$ receives a signal 
\begin{align*}
    s_i:=\bx_i^\top \btheta + \mathcal{N}(0,1).
\end{align*}
The noise that this signal introduces ensures a degree of privacy with respect to the user's data.\footnote{We set the noise variance to one to simplify the results. The analysis remains valid for any (potentially heterogeneous) level of noise.}

The data buyer aims to estimate a linear function of the user's data, $\by^\top \btheta$. Here, $\by \in \mathbb{R}^d $ with $ \|\by\|=1 $ denotes the data buyer's known characteristic vector{, where here and throughout the paper we use the two-norm}. If platform $i$ possesses the user's data (i.e., $a_i=1$), they present the data buyer with an option to acquire a noisy version of their data at a price $p_i$. Therefore, the \textbf{data buyer strategy profile is} $\bb \in \{0,1\}^K$, with $b_i=1$ denoting their acceptance of the platform $i$'s offer. The data buyer then receives a signal 
\begin{align*}
\tilde{s}_i := s_i + \mathcal{N}(0, \sigma_i^2) 
\end{align*}
for the price $p_i$, where $\sigma_i^2$ represents the variance of the noise that platform $i$ adds to the data they share with the data buyer. In light of the discussion on differential privacy in the introduction, adding Gaussian noise—known as the Gaussian mechanism—makes the user's data \emph{{locally} differentially private}. This means that a change in a user’s data has a limited impact on the distribution of the shared signal $\tilde{s}_i$, where the extent of the impact is controlled by the variance $\sigma_i^2$. This, in turn, limits how much information the signal $\tilde{s}_i$ reveals about the user’s data to the data buyer.\footnote{{More formally, a randomized mapping $\mathcal{M}(\cdot)$ is said to be $(\varepsilon,\delta)$-locally differentially private if
$\mathbb{P}(\mathcal{M}(s) \in W) \leq
\exp(\varepsilon) \mathbb{P}(\mathcal{M}(s') \in W) + \delta,$
for any $s, s'$ and any measurable set $W$. See \cite{duchi2016lecture} for further details and the characterization of the Gaussian mechanism.}}

Lastly, each platform $i \in [K]$ decides on the noise level to add to the user data before offering it to the data buyer, denoted by $\sigma_i \in \mathbb{R}_+$. We find it more convenient to work with the standard deviation of the noise $\sigma_i$ and assume, without loss of generality, it is positive rather than the variance $\sigma_i^2$. Each platform also decides whether to \emph{enter} the market and provide services to the user, denoted by $e_i \in\{0,1\}$. Therefore, \textbf{the platforms' strategy profile is} $(\bsigma, \be) \in \mathbb{R}_+^K \times \{0,1\}^K$.

\begin{figure}
  \centering
\includegraphics[width=0.9\textwidth]{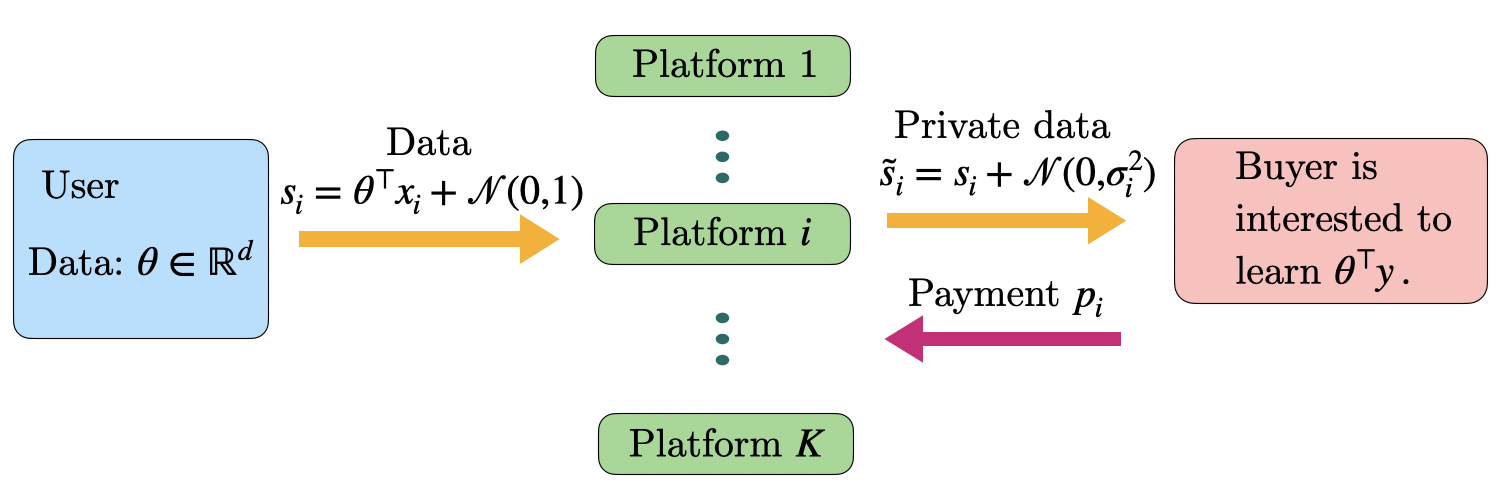}
    \caption{{A Summary of the Model.}}
    \label{fig:Model:Three_Layer}
\end{figure}

We set $\tilde{s}_i = \text{NA}$ when the data buyer does not obtain any information from platform $i$. This can occur either because the user has not shared their data with platform $i$ or because the data buyer declines the offer from platform $i$ or the platform has not provided the service to the user. In this case, $p_i$ is irrelevant, and we set it to zero. { A summary of the model is illustrated in \cref{fig:Model:Three_Layer}.}

Throughout the paper, we make the following assumption:
\begin{assumption} \label{assumption:substitute} 
Let \begin{align*}
    \gamma_i=\bx_i^T \by \text{ for all } i \in [K]
\end{align*}
denote the correlation between {platform} $i$'s data and the data buyer. Then, we assume
\begin{align*}
    \left(\bx_i-\gamma_i \by \right)^T\left( \bx_j-\gamma_j \by  \right)=0 \text{ for all } i\neq j \in [K].
\end{align*}
\end{assumption}
Note that, under this assumption, the data offered by different platforms are viewed as substitutes for each other from the data buyer's perspective. { As an example, suppose the data buyer is interested in inferring whether a household has children. The buyer collects data from both a streaming platform and an online retailer. The overlapping component of these two data sources such as evidence of child-oriented content consumption or purchases is precisely the information about whether the family has children. Once this common factor is taken into account, the remaining parts of the two data streams are largely unrelated, making the datasets substitutes rather than complements.}

However, if this assumption does not hold, a complementary effect could also be observed. To illustrate this, suppose we have $K=2$ platforms where $x_1$ is orthogonal to $y$, i.e., $\gamma_1= 0$, and also $x_2 = \gamma_2 y + (1-\gamma_2) x_1$. In this scenario, the data from platform 1 alone has no value to the data buyer, as $x_1^\top \theta$ provides no information about $\theta^\top y$. However, if the data buyer decides to purchase data from the second platform, then the data from the first platform could also become valuable. To understand why, observe that the data from the second platform is a convex combination of two components: $x_1^\top \theta$ and $y^\top \theta$. Therefore, if the data buyer acquires the data from the second platform, learning $x_1^\top \theta$ could assist in better distinguishing these two components. In other words, the data from the first platform is valuable only if accompanied by the data from the second platform. This scenario exemplifies the complementary effect we described earlier. Assumption \ref{assumption:substitute} excludes this possibility, allowing us to primarily focus on the substitution effect.
This assumption also implies that the correlation among data for different platforms is given by
\begin{equation*}
\bx_i^\top \bx_j = \gamma_i \gamma_j.   
\end{equation*}

\subsection{Utility functions}
To define the utility functions, we need to first quantify how much information is revealed by observing the signals. We do so by using the following definition, which captures the reduction in uncertainty following the observation of a set of signals. 
\begin{definition}[Revealed information]
\label{def:leakedinformation} Consider a random variable $Z$ drawn from the distribution $\pi(.)$. Let $\pi_\mathcal{S}(.)$ represent the posterior distribution of $Z$ upon observing a set of signals $\mathcal{S}$. The revealed information about $Z$ by observing $\mathcal{S}$ is quantified as the reduction in the mean-squared error of $Z$, i.e., 
\[
\mathcal{I}(Z ~|~\mathcal{S})=\mathbb{E}\left[ \left( Z -\mathbb{E}%
_{Z \sim \pi}\left[ Z \right] \right) ^{2}\right] -\mathbb{E}%
\left[ \left( Z -\mathbb{E}_{Z \sim \pi _{\mathcal{S}}}%
\left[ Z \right] \right) ^{2}\right] ,
\]
where the outer expectations are over the randomness in $Z$ and the signals.
\end{definition}
Measuring the revealed information, or information gain, through the variance of the posterior distribution has been used widely in economics, statistics, and learning theory \citep[see, e.g.,][]{breiman1984classification, bergemann2020economics,acemoglu2019too}.

Given this definition, the user's utility function is given by 
\begin{equation} \label{eqn:utility_user}
\mathcal{U}_{\text{user}}(\bsigma, \be, \ba, \bp, \bb):= 
\sum_{i=1}^K a_i  e_i \mathcal{I}_i
- \alpha ~ \mathcal{I}(\bsigma, \be, \ba, \bb),
\end{equation}
where the term
\begin{align*}
    \mathcal{I}_i \triangleq \mathcal{I}(\bx_i^\top \btheta  ~|~ s_i)
\end{align*}
is the user's revealed information to platform $i$ if the user shares her information (i.e., when $a_i=1$) and the platform is offering service to the user (i.e., when $e_i=1$).\footnote{{The user's utility is additive across platforms, reflecting environments where the services are not substitutes and the user can engage with multiple platforms simultaneously.}} The term
\begin{align*}
    \mathcal{I}(\bsigma, \be, \ba, \bb) \triangleq \mathcal{I} (\by^\top \btheta  ~|~  \bts),
\end{align*}
is the user's revealed information to the data buyer.\footnote{{The match between the user and the platform is captured by the term $x_i^T \theta$ and the user's gain is the platform's ability to learn (and act on) the match to deliver higher-quality service.}}  Here, $\bts$ denotes the vector $(\tilde{s}_i)_{i=1}^K$, which is the revealed information to the data buyer. Notice that the user's utility does not directly depend on $\bp$, but we retain this notation to highlight that through the data buyer's decision, the revealed information depends on $\bp$. Also, notice that in finding the revealed information to the data buyer, all that matters is the set $\{\sigma_i~:~ \text{ s.t. } a_i=b_i=e_i=1\}$. Therefore, we also find it convenient to define
\begin{align}\label{def:shorthandInformation}
    \mathcal{I}(\bsigma_{S}) \text{ for } \bsigma_S=(\sigma_i~:~ i \in S ) \text{ for any } S \subseteq [K],
\end{align}
where we may drop the subindex $S$ whenever the indices of the elements of $\bsigma$ are clear from the context.

The first term in \eqref{eqn:utility_user} captures the user's gain from the service offered by each platform. Notice that this term is present if the platform has entered the market (i.e., $e_i=1$) and if the user shares with the platform (i.e., $a_i=1$). The second term represents the user's privacy loss from the sharing of their data with the data buyer. Additionally, $\alpha$ indicates the user's relative importance on these two opposing terms.  

For each $i \in  [K]$, platform $i$'s utility is given by
\begin{equation}\label{eqn:utility_platform_i}
\mathcal{U}_{i}(\bsigma, \be, \ba, \bp, \bb):= \begin{cases}
a_i \mathcal{I}_i + b_i p_i - c_i\quad & e_i=1\\
0 \quad &e_i=0.
\end{cases} 
\end{equation}
When the platform has not entered the market (i.e., $e_i=0$), the platform's utility is zero. When the platform has entered the market (i.e., $e_i=1$), the second term $b_i p_i$ represents the platform's revenue from selling the user's data to the data buyer. The first term $a_i \mathcal{I}_i$, similar to the user's utility, reflects the quality of service the platform provides (if the user ends up utilizing their service). This term is included to capture the fact that a better service by the platform also enhances the platform's utility. For example, it could increase ad revenue or expand the user base. Lastly, the cost $c_i \in \mathbb{R}_+$ reflects the per-user cost incurred by the platform for providing the service.

Finally, the data buyer's utility is given by
\begin{equation}\label{eqn:utility_buyer}
\mathcal{U}_{\text{buyer}}(\bsigma, \be, \ba, \bp, \bb):= 
\beta ~ \mathcal{I}(\bsigma, \be, \ba, \bb)  - \sum_{i=1}^K b_i p_i,
\end{equation}
which is the data buyer's gain from learning the user's information minus the amount they pay to all the platforms to purchase their data. Here, $\beta \in [0,1]$ is some constant that captures the importance of the revealed information compared to the payments in the data buyer's utility. 

\begin{figure}
  \centering
\includegraphics[width=0.8\textwidth]{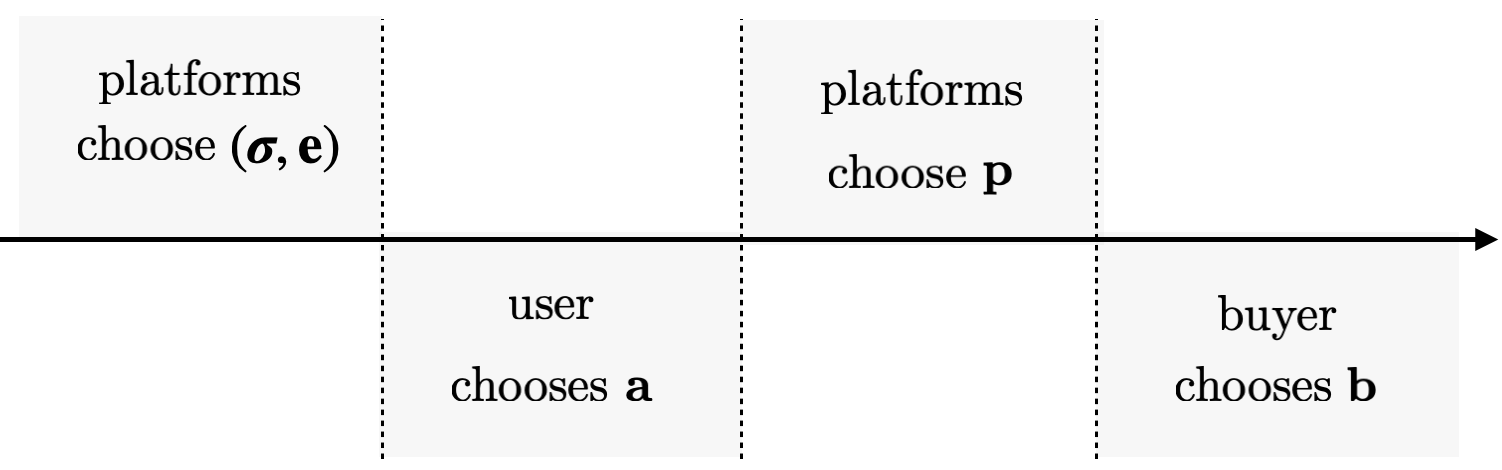}
    \caption{The timing of the decisions.}
    \label{fig:timeline}
\end{figure}

We refer to the above game among the platforms, user, and buyer of data as a \emph{three-layer data market} with parameters $(\alpha, \beta, \bc, \bgamma)$. Here is the timing of the game (see Figure \ref{fig:timeline}): 
\begin{enumerate}
    \item In the first stage, all platforms simultaneously choose $(\bsigma, \be) \in \mathbb{R}_+^K \times \{0,1\}^K$ that specifies their noise levels and whether they want to enter the market to provide services to the user.
    \item In the second stage, the user chooses $\ba \in \{0,1\}^K$ that specifies which platforms to join and share data with.
    \item In the third stage, the platforms simultaneously choose $\bp \in \mathbb{R}^K$ that determines their offered prices.
    \item In the fourth stage, the data buyer chooses $\bb \in \{0,1\}^K$ that determines whether they accept or decline each offer made by the platforms.
\end{enumerate}

{
\begin{remark}
    \textup{In our baseline model, we assume the utility of all players to be linear in the leaked information. In Appendix \ref{section:extension_nonlinear}, we establish that our main results continue to hold more broadly. Specifically, we let the user utility be\footnote{We retain a linear form for the first $\mathcal{I}_i$ because this is a constant term and applying a function to it does not change the analysis.} 
\begin{equation} \label{eqn:utility_user:extend}
\mathcal{U}_{\text{user}}(\bsigma, \be, \ba, \bp, \bb):= 
\sum_{i=1}^K a_i  e_i \mathcal{I}_i
- \alpha ~ H_u\left(\mathcal{I}(\bsigma, \be, \ba, \bb) \right),
\end{equation}
and the 
data buyer's utility be
\begin{equation}\label{eqn:utility_buyer:extend}
\mathcal{U}_{\text{buyer}}(\bsigma, \be, \ba, \bp, \bb):= 
\beta ~ H_b\left(\mathcal{I}(\bsigma, \be, \ba, \bb) \right) - \sum_{i=1}^K b_i p_i.
\end{equation}
We then prove that our results continue to hold if the functions $H_u(\cdot)$ and $H_b(\cdot)$ belong to the following family of functions:
\begin{align*}
    \mathcal{H}=\{H:[0,1] \mapsto [0,1]~:~ \inf_{x \in [0,1]} H'(x) >0, ~H(\cdot) \text{ is concave}, ~ \text{ and } H(0)=0\}.
\end{align*}
    }
\end{remark}

}

\subsection{Revealed Information Properties}
Before analyzing this game, we state some properties of our notion of revealed information. We recall that the leaked information to the data buyer depends on $\be$, $\ba$, and $\bb$ through the set of platforms that share user data with the data buyer and the noise level of each of them.  Therefore, we make use of the shorthand notation for the leaked information given in \eqref{def:shorthandInformation}. In what follows, we also consider the \emph{lattice} $(\mathcal{L}, \le)$, where
\begin{align}\label{eq:def:Lattice}
    \mathcal{L}:=\{\ell ~|~ \ell \in\{0,1\}^K \} \text{ and }
    \ell \le \ell' \text{ if } \ell_i \le \ell'_i \text{ for all } i \in [K]. 
\end{align}

The revealed information admits the following closed-form expression.
\begin{lemma}\label{Lem:revealed:closedform}
For any $i \in [K]$, we have $\mathcal{I}_i=1/2$. Moreover, for any $\bsigma$ and $S \subseteq [K]$ we have
\begin{align*}
    \mathcal{I}(\bsigma_S)= \mathbf{m}^T M^{-1} \mathbf{m},
\end{align*}
where $m_i=\gamma_i$, $[M]_{ii}=2+\sigma_i^2$, and $[M]_{ij}=\gamma_i \gamma_j$ for all $i , j \in S$ such that $i \neq j$.
\end{lemma}
\cref{Lem:revealed:closedform} follows by using the properties of the multivariate normal distribution.\footnote{{For simplicity, we assumed that the user adds the same noise level with variance one to their data before sharing it with any platform, resulting in $\mathcal{I}_i=1/2$ for all $ i \in [K]$. Our analysis continues to hold for any
heterogeneous noise levels.}}

Importantly, \cref{Lem:revealed:closedform} enables us to obtain some structural properties of the revealed information, as we now demonstrate.

\begin{lemma}[monotonicity]\label{Lem:monotone}
 The revealed information is increasing in the set of platforms that share the user data with the data buyer and is decreasing in $\bsigma$; i.e., for any $\bsigma$ and $S \subseteq [K]$, $\mathcal{I}(\bsigma_S)$ is increasing in $S$ and decreasing in $\bsigma$. 
\end{lemma}
Lemma \ref{Lem:monotone} simply states that more data sharing with the data buyer increases its revealed information. Moreover, data with a lower noise level increases the revealed information to the data buyer. 
\begin{lemma}[submodularity in actions]\label{Lem:submodular}
${}^{}$
For a given vector of noise levels $\bsigma$, the revealed information is submodular in the set of platforms that share the user data with the data buyer; i.e., for any $\bsigma$, $S \subseteq T \subseteq [K]$, and $i \not\in T$, we have 
\begin{align*}
    \mathcal{I}((\sigma_i, \bsigma_S))-\mathcal{I}(\bsigma_S) \ge \mathcal{I}((\sigma_i, \bsigma_T))-\mathcal{I}(\bsigma_T). 
\end{align*}
\end{lemma}
\cref{Lem:submodular} states the intuitively appealing result that as more data is shared with the data buyer, the marginal increase in the revealed information from one more unit of shared data becomes smaller.  
\begin{lemma}[submodularity in noise variance]\label{Lem:submodular:sigma}
${}^{}$
For any $\bsigma$, $S \subseteq [K]$, $i \not\in S$, and $j \in S$
\begin{align*}
    \mathcal{I}((\sigma_i, \bsigma_S))-\mathcal{I}(\bsigma_S) 
\end{align*}
is decreasing in $\sigma_i$ and increasing in $\sigma_j$. 
\end{lemma}
Lemma \ref{Lem:submodular:sigma} states that the data of platform $i$ reveals a relatively larger amount of information for smaller values of $\sigma_i$ and for larger values of $\sigma_j$. To understand the former, let us compare the case of $\sigma_i \approx 0$ and $\sigma_i \to \infty$. When  $\sigma_i \approx 0$, the data of platform $i$ reveals a lot, which helps the data buyer learn better. However, when $\sigma_i \to \infty$, the data of platform $i$ does not contain any information, and therefore, the relative gain in the revealed information is $\approx 0$. To understand the latter, let us again compare the case of $\sigma_j \approx 0$ and $\sigma_j \to \infty$. When  $\sigma_j \approx 0$, the data of platform $j$ reveals a lot; therefore, the marginal increase in the revealed information of platform $i$ is small. However, when $\sigma_j \to \infty$, the data of platform $j$ does not contain any information, and therefore, the relative increase in the revealed information from platform $i$'s data is much larger.

{The lemmas above, which establish various properties of the revealed information, allow us to characterize the equilibrium properties in the next section. Although the derivation of these lemmas relies on \Cref{ass:1} and the definition of revealed information, one can relax that assumption or adopt an alternative notion of information gain,
and our results on the equilibrium properties would continue to hold as long as the aforementioned properties remain valid.}
 
\section{Preliminary Equilibrium  Characterization}\label{Sec:Preliminary}
We adopt subgame perfect equilibrium as our solution concept, which means that the strategy profile of all players is a Nash equilibrium of every subgame of the game. We next formally define and characterize the subgame perfect equilibrium in our setting, starting from the last stage.

\paragraph*{Buyer equilibrium:} For a given $(\bsigma, \be, \ba, \bp)$, a data buyer profile $\bb^{\mathrm{E}}$ is an equilibrium if it achieves the data buyer's highest utility, i.e., 
\begin{align*}
    \bb^{\mathrm{E}} \in \argmax_{\bb \in \{0,1\}^K} \mathcal{U}_{\text{buyer}}(\bsigma, \be, \ba, \bp, \bb).
\end{align*}
Given that there is a finite set of $2^n$ possibilities, one of them achieves the maximum data buyer's utility, and therefore, an equilibrium buyer profile strategy always exists. We let $\mathcal{B}^{\mathrm{E}}(\bsigma, \be, \ba, \bp)$ be the set of such buyer equilibria. 

\paragraph*{Price equilibrium:} For a given $(\bsigma, \be, \ba)$, a pricing strategy by the platforms $\bp^{\mathrm{E}}$ is an equilibrium if no platform has a profitable deviation (taking the data buyer's updated decision into account):
\begin{align*}
& \mathcal{U}_i(\bsigma, \be, \ba, \bp^{\mathrm{E}}, \bb^{\mathrm{E}}) \ge \mathcal{U}_i(\bsigma, \be, \ba, (p_i,\bp^{\mathrm{E}}_{-i}), \bb) \\
& \text{ for all } i \in [K], \bb^{\mathrm{E}} \in \mathcal{B}^{\mathrm{E}}(\bsigma, \be, \ba, \bp^{\mathrm{E}}), p_i, \bb\in \mathcal{B}^{\mathrm{E}}(\bsigma, \be, \ba, (p_i,\bp^{\mathrm{E}}_{-i})).
\end{align*}
Before proceeding with the rest of the game, we establish that the data buyer and price equilibria admit a simple characterization, as stated next. 
\begin{proposition}
\label{Pro:Equilibirum:Buyer:Price}
For a given $(\bsigma, \be, \ba)$, there exists a unique price equilibrium given by 
\begin{align*}
    p_i^*= \beta \left( \mathcal{I}(\bsigma, \be, \ba, \mathbf{1}) - \mathcal{I}(\bsigma, \be, \ba, (b_i=0,\mathbf{1}_{-i})) \right) \text{ for all } i \in [K]
\end{align*}
and the corresponding unique buyer equilibrium is $b_i=1$ for all $i \in [K]$, where $\mathbf{1}_{-i}$ is the vector of all ones for all $j \in [K] \setminus \{i\}$. 
\end{proposition}
{ To see why this result holds, first note that for any platform $i$ that has entered the market and with which the user has shared data (i.e., $e_i = a_i = 1$), we have $b_i = 1$ in equilibrium. This is because if that were not the case, the platform would have a profitable deviation by setting a very low but positive price for its data to incentivize the buyer to purchase it. This positive price exists because, if the buyer obtains the platform’s data, their utility increases; thus, any price below that positive utility gain multiplied by $\beta$ would still incentivize the buyer to buy the data.

Next, knowing that all these $b_i$'s are equal to one in equilibrium, we claim that the prices given in the statement of \cref{Pro:Equilibirum:Buyer:Price} constitute an equilibrium. The proof is straightforward: no platform has an incentive to decrease this price (since the buyer would still be willing to pay more), nor to increase it (since the buyer would no longer purchase from them). The proof of uniqueness is provided in the Appendix.}

\cref{Pro:Equilibirum:Buyer:Price} characterizes the equilibrium of the third and fourth stages of the game. Therefore, it remains to characterize the user sharing profile and the platform's noise level and entry decisions in equilibrium, taking into account the equilibrium of the subsequent stages that we have already identified.  
\paragraph*{User equilibrium:} For a given $(\bsigma, \be)$, a user sharing profile $\mathbf{a}^{\mathrm{E}}$ is an equilibrium if it achieves the user's highest utility, i.e., 
\begin{align*}
    \ba^{\mathrm{E}} \in \argmax_{\ba \in \{0,1\}^K} \mathcal{U}_{\text{user}}(\bsigma, \be, \ba, \bp^*, \mathbf{1}).
\end{align*}
Again, given that there are $2^n$ possibilities, one of them achieves the maximum user's utility, and therefore, an equilibrium sharing profile strategy always exists. We let $\mathcal{A}^{\mathrm{E}}(\bsigma, \be)$ be the set of such user equilibria. This set has the following important lattice structure that we use in the rest of the analysis.
\begin{lemma}\label{Lem:equilibrium:lattice}
    For a given $(\bsigma, \be)$, the set of user equilibria $\mathcal{A}^{\mathrm{E}}(\bsigma, \be)$ is a sublattice of the lattice $(\mathcal{L}, \le)$ (defined in \eqref{eq:def:Lattice}). Therefore, it has a maximum and a minimum. 
\end{lemma}
For a platform strategy profile $(\bsigma, \be)$, we select the platforms' Pareto-optimal one if there are multiple user equilibria. Using \cref{Lem:equilibrium:lattice}, such a user equilibrium exists and is the one that shares with the highest number of platforms.\footnote{We can view this as Stackelberg Nash equilibrium in which because the platforms move first, we break the ties in their favor. Similar concepts have appeared in the literature with different terms, such as sender-preferred equilibrium in Bayesian persuasion \citep{kamenica2011bayesian}.}

\paragraph*{Platform equilibrium:} A noise level and entry strategy $(\bsigma^{\mathrm{E}}, \be^{\mathrm{E}})$ is an equilibrium if no platform has a profitable deviation, i.e.,
\begin{align*}
    \mathcal{U}_i(\bsigma^{\mathrm{E}}, \be^{\mathrm{E}}, \ba^{\mathrm{E}}, \bp^*, \mathbf{1}) \ge \mathcal{U}_i((\sigma_i,\bsigma_{-i}^{\mathrm{E}}), (e_i, \be_{-i}^{\mathrm{E}}), \ba, \bp^*, \mathbf{1})
\end{align*}
for all  $i \in [K], \ba^{\mathrm{E}} \in \mathcal{A}^{\mathrm{E}}(\bsigma^{\mathrm{E}}, \be^{\mathrm{E}}), \ba \in \mathcal{A}^{\mathrm{E}}((\sigma_i,\bsigma_{-i}^{\mathrm{E}}), (e_i, \be_{-i}^{\mathrm{E}}))$.

We refer to a platform's noise level and entry decision as the platform's decision and to the corresponding equilibrium as the platforms' equilibrium. The following characterizes the set of possible platforms' equilibria. 
\begin{proposition}
\label{Pro:Equilibirum:User:Noise}
For any platforms' equilibrium strategy $(\bsigma, \be)$, we must have 
\begin{align*}
    \mathcal{A}^{\mathrm{E}}(\bsigma, \be)=\{(a_i=1~:~ \text{ for all } i \text{ s.t. } e_i=1))\},
\end{align*}
that is, the user must find it optimal to share with all platforms that have entered the market. 
\end{proposition}
This proposition follows by noting that a platform that has entered the market can always increase its noise level so that the user finds it optimal to share with the platform.

Motivated by \cref{Pro:Equilibirum:User:Noise} and \cref{Lem:equilibrium:lattice}, we conclude that for any set of platforms that have entered the market (captured by $\be \in \{0,1\}^K$), the set of possible noise levels in equilibrium is $\Sigma_{\mathbf{1}, \be}$, where we have
\begin{align} \label{eqn:regions}
    \Sigma_{\mathbf{\ba}, \be}:=\{(\sigma_i~:~ i \text{ s.t. } e_i=1 ) ~:~ \ba \text{ is the highest element of } \mathcal{A}^{\mathrm{E}}(\bsigma, \be) \} \text{ for all } \ba \in \{0,1\}^K.
\end{align}
We are interested in characterizing $\Sigma_{\mathbf{1}, \be}$ for any $\be$ and finding the equilibrium that belongs to one of these sets for an equilibrium choice of $\be$. 

Throughout the paper, we focus on the case in which the user asks for some level of privacy in order to share their data. This means that if the noise levels are all zero, then the user does not share with the platforms. More formally, we make the following assumption:
\begin{assumption} \label{assumption:all_regions}
The weight of privacy on the user utility, i.e., $\alpha$, is large enough such that, for any $\be \in \{0,1\}^K$, the point $(\sigma_i=0~:~ i \text{ s.t. } e_i=1)$ belongs to the set $\Sigma_{\mathbf{0}, \be}$.      
\end{assumption}
We will further clarify this assumption's role in the next section. 
\section{Equilibrium Characterization with $K=2$ Platforms}\label{sec:Equilibrium:K=2}
We begin our analysis by considering a setting with $K=2$ platforms. We then show that the main insights from this analysis extend to the general setting with $K>2$ platforms. 

\begin{figure}
\centering
\begin{subfigure}{.5\textwidth}
  \centering
  \includegraphics[width=\linewidth]{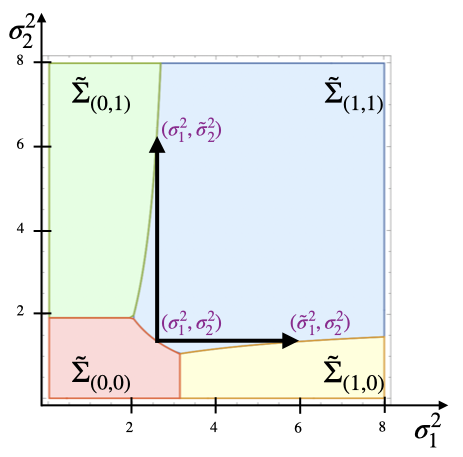}
  \caption{}
  \label{fig:RegionsK=2}
\end{subfigure}
\begin{subfigure}{.5\textwidth}
  \centering
  \includegraphics[width=\linewidth]{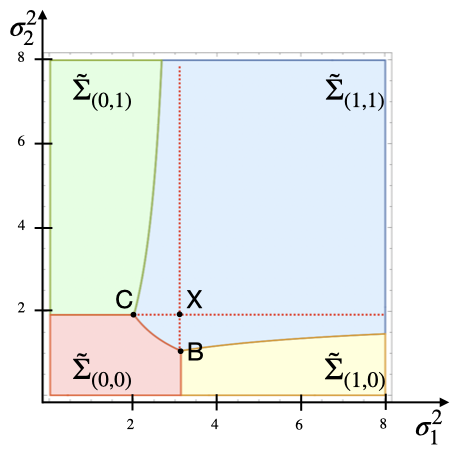}
  \caption{}
  \label{fig:OneParticipate} 
\end{subfigure}
\caption{Four possible sets $\tilde{\Sigma}_{\ba}$ for $\ba \in \{0,1\}^2$ for a setting with $K=2$ platforms, $\gamma_1=.8$, and $\gamma_2=.7$.}
\label{fig:RegionsK=2_all}
\end{figure}

Depending on which platforms enter the market, we have four cases:
\paragraph*{Case 1 where both platforms enter the market:} In this case, the user's utility if they share their data with both platforms is given by\footnote{With a slight abuse of notation, we drop the other dependencies of the user utility here.} 
\begin{equation}
\mathcal{U}_{\text{user}}(\sigma_1, \sigma_2) := 1 - \alpha \mathcal{I}(\sigma_1, \sigma_2),     
\end{equation}
and the user's utility if they share their data with platform $i \in \{1,2\}$ only is equal to
\begin{equation}
\mathcal{U}_{\text{user}}(\sigma_i) := \frac{1}{2} - \alpha \mathcal{I}(\sigma_i).   
\end{equation}
Recall the definition of $\Sigma_{\bm{a}, \bm{e}}$ with $\bm{e}=(1,1)$ in \eqref{eqn:regions}. To simplify the notation, we drop the $\bm{e}$ and use $\tilde{\Sigma}_{\bm{a}}$. In particular, 
\begin{align*}
    \tilde{\Sigma}_{(1,1)}:= \{(\sigma_1, \sigma_2)~:~ \mathcal{U}_{\text{user}}(\sigma_1, \sigma_2) = \max\left(\mathcal{U}_{\text{user}}(\sigma_1, \sigma_2), \mathcal{U}_{\text{user}}(\sigma_1), \mathcal{U}_{\text{user}}(\sigma_2), 0 \right)\}.
\end{align*}
Similarly, $\tilde{\Sigma}_{(1,0)}$ and $\tilde{\Sigma}_{(0,1)}$ denote the cases where $\mathcal{U}_{\text{user}}(\sigma_1)$ and $\mathcal{U}_{\text{user}}(\sigma_2)$, respectively, have the highest value. Finally,
\begin{align*}
     \tilde{\Sigma}_{(0,0)}:= \{(\sigma_1, \sigma_2)~:~ 0=\max\left(\mathcal{U}_{\text{user}}(\sigma_1, \sigma_2), \mathcal{U}_{\text{user}}(\sigma_1), \mathcal{U}_{\text{user}}(\sigma_2), 0 \right)\}.
\end{align*}
Notice that we operate under Assumption \ref{assumption:all_regions}, which implies that $\tsnn$ is non-empty. In particular, for $K=2$, this assumption translates to
\begin{equation} \label{eqn:assumption_K=2}
\alpha > \frac{4-\gamma_1^2 \gamma_2^2}{2\left( \gamma_1^2+\gamma_2^2-\gamma_1^2\gamma_2^2 \right)}.    
\end{equation}
Figure \ref{fig:RegionsK=2} depicts the sets $\tsyy$, $\tsny$, $\tsyn$, and $\tsnn$ as a function of the noise variance that the two platforms add to the data, i.e., $(\sigma_1^2, \sigma_2^2)$, for parameters $\gamma_1=0.8$ and $\gamma_2=0.7$. 

Using \cref{Pro:Equilibirum:User:Noise}, the noise level equilibrium is in the set $\tsyy$. In fact, we argue that it must belong to $\tsyy \cap \tsnn$ because, otherwise, one of the platforms can increase its utility by decreasing its noise level such that the user still shares with it. 




Now, for any $(\sigma_1, \sigma_2) \in \tsyy \cap \tsnn$, the only possible deviation for platform $1$ is to increase $\sigma_1$ to $\tilde{\sigma}_1$ to move to the point $(\tilde{\sigma}_1, \sigma_2)$ which resides in $\tsyn$, exactly next to the boundary with $\tsyy$ (see Figure \ref{fig:RegionsK=2}). At first glance, it might look like this can never be a deviation because it reduces the revealed information to the buyer. However, notice that, by this action, platform $1$ pushes platform $2$ out of the market because $(\tilde{\sigma}_1, \sigma_2)$ is in the yellow region $\tsyn$ in which the user does not share their data with the second platform. Therefore, it is possible that the first platform's payment increases, as they are now the only platform with access to the data. Note that this is the only deviation that we need to consider. Because if it ends up not being a deviation, it would imply that a further increase of $\sigma_1$ into the yellow region is also not profitable. Also, it is evident that increasing $\sigma_1$ but staying in the $\tsyy$ region or decreasing it and going into the $\tsnn$ region in which the user shares no data are not profitable deviations. Considering these deviations results in the following.

\begin{lemma} \label{lemma:alpha_1.88}
Suppose Assumptions \ref{assumption:substitute} and \ref{assumption:all_regions} hold. Assuming both platforms have entered the market, there exists an equilibrium noise level $(\sigma_1, \sigma_2) \in \tsyy \cap \tsnn$ if and only if $\alpha \geq \bar{\alpha} \approx 1.884$. 
 In particular, for $\alpha \geq \bar{\alpha}$, $(\sigma_1, \sigma_2)$ is an equilibrium where 
 \begin{equation} \label{eqn:equilibrium_K=2}
 \frac{2+\sigma_1^2}{\gamma_1^2}  
 = \frac{2+\sigma_2^2}{\gamma_2^2}  
 = 2\alpha-1.
 \end{equation}
\end{lemma}
\cref{lemma:alpha_1.88} is under the assumption that both platforms enter the market. The final point to check is whether any platform has an incentive to enter the market. 
\begin{proposition} \label{eqn:proposition_equilibrium_pp}
Suppose Assumptions \ref{assumption:substitute} and \ref{assumption:all_regions} hold. Also, recall $\bar{\alpha} \approx 1.88$ from \cref{lemma:alpha_1.88}. If an equilibrium exists in which both platforms enter the market, then  
\begin{equation} \label{eqn:necessary_beta_K=2}
\alpha \geq \bar{\alpha} \quad \text{and} \quad
\beta \geq 2\alpha~(\max\{c_1,c_2\}-\frac{1}{2}).
\end{equation}
Conversely, for any $\alpha \geq \bar{\alpha}$ and any $\beta \geq \bar{\beta}(\alpha)$ such an equilibrium in which both platforms enter the market exists, where
\begin{equation} \label{eqn:sufficient_beta_K=2}
\bar{\beta}(\alpha) \triangleq (2 + \frac{1}{\alpha-1})~\alpha~(\max\{c_1,c_2\}-\frac{1}{2}).
\end{equation}
\end{proposition}
Both parts of the result provide a lower bound on $\beta$, with the difference that the first part establishes a necessary condition for an equilibrium to exist, while the second part presents a sufficient condition. Note that the ratio between these two lower bounds converges to one as $\alpha$ grows. However, we next highlight a subtle difference between the two results. Note that, for any value $\xi$ greater than $2~(\max{c_1,c_2}-\frac{1}{2})$, we can choose $\alpha$ and $\beta$ sufficiently large such that an equilibrium exists for these parameters and the ratio of $\beta$ over $\alpha$ is less than or equal to $\xi$. In other words, we have 
\begin{equation*}
 \inf \left \{ \frac{\beta}{\alpha} ~\Big \vert ~ \alpha, \beta \text{ s.t. an equilibrium exists} \right \} = 2 \left (\max\{c_1,c_2\}-\frac{1}{2} \right ).
\end{equation*}
This, in fact, is a direct consequence of the second part of \cref{eqn:proposition_equilibrium_pp} along with the fact that $\bar{\beta}$ converges to $2~(\max\{c_1,c_2\}-\frac{1}{2})$ as $\alpha$ grows. 

However, the form of the infimum above does not imply that any $\alpha$ and $\beta$ satisfying the condition \eqref{eqn:necessary_beta_K=2} will necessarily result in an equilibrium. In fact, as our proof establishes, to have an equilibrium for $\alpha = \bar{\alpha}$, we must have $\beta \geq \bar{\beta}(\bar{\alpha})$. In other words, we have
\begin{align*}
\sup_{\alpha \geq \bar{\alpha}} \inf \left \{ \frac{\beta}{\alpha} ~\Big \vert ~ \beta \text{ s.t. an equilibrium exists with the given } \alpha \right \} = \bar{\beta}(\bar{\alpha}).
\end{align*}
\paragraph*{Cases 2 and 3 where only one platform enters the market:} Suppose only platform $1$ enters the market at some equilibrium. In this case, the first platform wants to choose the noise level $\sigma_1$ as small as possible but also large enough so that user shares their data. Therefore, the platform chooses $\sigma_1$ at the value that makes the user indifferent between sharing and not sharing (recall that we assumed the user breaks the ties in favor of the platforms), i.e.,
\begin{equation} \label{eqn:one_user_participate}
\mathcal{U}_\text{user}(\sigma_1) = \frac{1}{2} - \alpha \mathcal{I}(\sigma_1) = 0,     
\end{equation}
which implies $\sigma_1^2 = 2\alpha \gamma_1^2 - 2$. Notice that this is equivalent to the values of $\sigma_1$ corresponding to points $B$ and $X$ in Figure \ref{fig:OneParticipate}.  

Now, for this noise level to be an equilibrium, we need to verify two potential deviations: (i) platform 1 has no incentive to abstain from participating, i.e., their utility should be nonnegative, and (ii) platform 2 has no profitable deviation by entering the market. Using \eqref{eqn:one_user_participate}, the first condition simply implies
\begin{equation} \label{eqn:one_participate_beta_lowerbound}
\frac{1}{2} - c_1 + \beta \frac{1}{2\alpha} \geq 0.
\end{equation}
For the second deviation check, we argue that it suffices to ensure the second platform's utility at point $B$ in Figure \ref{fig:OneParticipate} is nonnegative.

Let us denote the noise levels corresponding to point $B$ as $(\sigma_1^B, \sigma_2^B)$. To understand why we only need to check the point $B$, consider that, given platform 1's chosen noise level (equal to $\sigma_1^B$), if the second platform enters the market and selects a certain noise variance, it will result in a point on the line $BX$. The smallest noise level for the second platform that motivates the user to share their data with them (and thus has the potential to be profitable) is $\sigma_2^B$. One might argue that the second platform could opt for a very large noise level so that the vector of noise levels falls into the green region $\tsny$, implying that only the second platform receives the data. However, this is not feasible, as the line $BX$ stays within the blue region $\tsyy$ for any value of $\sigma_2$ (and in fact, it becomes tangent to the border of $\tsny$ and $\tsyy$ as $\sigma_2$ approaches infinity). To confirm this, we need to demonstrate that
\begin{equation*}
\mathcal{U}_{\text{user}}(\sigma_1^B, \sigma_2) = 1 - \alpha \mathcal{I}(\sigma_1^B, \sigma_2) \geq \frac{1}{2} - \alpha \mathcal{I}(\sigma_2) = \mathcal{U}_{\text{user}}(\sigma_2),
\end{equation*}
holds for any $\sigma_2$. This is indeed the case, given that due to the submodularity of the revealed information, $\mathcal{I}(\sigma_1^B, \sigma_2) - \mathcal{I}(\sigma_2)$ decreases with increasing $\sigma_2$ and equals $\mathcal{I}(\sigma_1^B) = \frac{1}{2\alpha}$ when $\sigma_2 = \infty$.

Now, checking the second platform's utility at point $B$ along with \eqref{eqn:one_participate_beta_lowerbound} yields the following result.
\begin{proposition}\label{proposition_equilibrium_pn}
Suppose Assumptions \ref{assumption:substitute} and \ref{assumption:all_regions} hold. For any $i \in \{1,2\}$, there exists an equilibrium in which only platform $i$ enters the market if and only if $c_i \leq c_{-i}$ and 
\begin{equation*}
\beta \in \left[ 2\alpha(c_i-\frac{1}{2}), 2\alpha(c_{-i}-\frac{1}{2}) \right]. 
\end{equation*}
\end{proposition}
\paragraph*{Case 4 where no platform enters the market:} Finally, we ask whether there is an equilibrium in which no platform enters the market. In such a scenario, we need to ensure that no platform has a profitable deviation by entering the market and choosing some noise level. For instance, if platform $1$ wants to deviate, the smallest noise level that they would choose would be $\sigma_1^B$ as it makes the user indifferent between sharing and not sharing their data with them. In this case, the platform's utility, as derived in \eqref{eqn:one_participate_beta_lowerbound}, is given by
\begin{equation*}
 \frac{1}{2} - c_1 + \beta \frac{1}{\alpha},   
\end{equation*}
which is nonpositive when $\beta \leq 2\alpha (c_1 - \frac{1}{2})$. Making the same argument for the second platform's deviation, we obtain the following result. 
\begin{proposition} \label{proposition:equilibrium_nn}
Suppose Assumptions \ref{assumption:substitute} and \ref{assumption:all_regions} hold. There exists an equilibrium in which neither of the platforms enters the market if and only if
\begin{equation*}
\beta \leq 2\alpha \left ( \min\{c_1, c_2\} - \frac{1}{2} \right ).    
\end{equation*}
\end{proposition}
\subsection{Insights from the case with $K=2$}
We conclude this section by highlighting a few implications and insights of our results for the case \( K=2 \), which, as we will see, also extend to the general case. Note that the term \( \beta \) determines the value the buyer places on the user's data. The higher \( \beta \) is, the more money the platform can earn by selling the user's data.

Our results indicate that there are, at most, three main regimes for \( \beta \), and the market equilibrium depending on the interval \( \beta \) falls into. First, as established in \cref{proposition:equilibrium_nn}, when \( \beta \) is small, the platform's monetary gain from selling data is limited, so platforms may decide not to provide any service as they cannot fully compensate for their service costs. This regime would not exist if at least one of the platforms operates at a \textit{low cost}, meaning its gain from the user's data already compensates for the cost of service (this corresponds to the case \( c_i \leq 1/2 \) for some \( i \)).

On the other hand, as shown in \cref{eqn:proposition_equilibrium_pp}, when \( \beta \) is large enough and platforms can charge a high price for the user's data, both platforms would enter the market. Finally, there is an intermediary regime for \( \beta \) where only the platform with the lower cost enters the market, and the platform with a higher service cost opts out.

Interestingly, the user's utility in all these cases would be equal to zero. However, the buyer's utility could be positive when both platforms enter the market. As we establish next, these observations continue to hold in the general case with $K>2$ platforms. 
\section{Equilibrium Characterization with $K>2$ Platforms}\label{sec:equilibrium:K>2}
This section shows how our equilibrium characterization extends to a general setting with any $K>2$ number of platforms. In what follows, we make use of the following definition.

\begin{definition}[High- and low-cost platforms]\label{definition:highlowcosts}
    We say a platform $i \in [K]$ is of high cost if $c_i > \frac{1}{2}$ and is of low cost if $c_i < \frac{1}{2}$.
\end{definition}
The comparison of the cost to $1/2$ in the above definition comes from the fact that the leaked information to any platform is $\mathcal{I}_i=1/2$ and therefore, high-cost platforms enter the market only if the payment they receive from the buyer is large enough. Conversely, low-cost platforms enter the market irrespective of the payment they receive from the buyer. We can view low-cost platforms as those that do not necessarily need additional payments from the data buyer to enter the market and provide services. These platforms might include large ones with small operating costs or those that do not rely solely on selling data to generate profit. On the other hand, high-cost platforms are the opposite; they would require some payment from the data buyer to be able to provide services.


\begin{theorem}\label{thm:equilibirum:K>2}
Suppose Assumptions \ref{assumption:substitute} and \ref{assumption:all_regions} hold. There exists $\bar{\alpha}$, $\bar{\beta}$, and $\underline{\beta}$ such that for $\alpha \ge \bar{\alpha}$, we have the following:
\begin{itemize}
    \item If $\beta \ge \bar{\beta}$, an equilibrium of the game exists. Moreover, in any equilibrium, all platforms enter the market, the user utility is zero, and the buyer purchases the data of all platforms and has a strictly positive utility.
    \item If $\beta \le \underline{\beta}$, an equilibirum of the game exists. Moreover, in any equilibrium, only low-cost platforms enter the market, the user utility is zero, and the buyer purchases the data of low-cost platforms and has a strictly positive utility only if there is more than one low-cost platform.
\end{itemize}
\end{theorem}
Let us explain the qualifiers of \cref{thm:equilibirum:K>2} and then explain its intuition. First, notice that, similar to the setting with $K=2$ platforms, for small enough $\alpha$ and also for intermediary values of $\beta$, a pure-strategy equilibrium may not exist. Therefore, we focus on large enough $\alpha$ and also avoid intermediary values of $\beta$. In the proof of \cref{thm:equilibirum:K>2}, we have provided explicit expressions for $\bar{\alpha}$, $\bar{\beta}$, and $\underline{\beta}$. {In particular, $\bar{\alpha}=(K+1)^2/8$ depends only on $K$ while $\underline{\beta}$ and $\bar{\beta}$ depend on the model parameters $\{(c_i, \gamma_i)\}_{i\in [K]}$.}

To gain the intuition of the first part of \cref{thm:equilibirum:K>2}, notice that for large enough $\beta$ the data buyer gains a lot when the user's data is revealed to them and therefore is willing to pay a high price for the user data.  This, in turn, means that even high-cost platforms find it optimal to enter the market. As established in \cref{Pro:Equilibirum:Buyer:Price} and \cref{Pro:Equilibirum:User:Noise}, in equilibrium, the user shares with all platforms, and the data buyer purchases from all platforms. To see that the user utility becomes zero in equilibrium, suppose the contrary, that the user utility is strictly positive. This means that the equilibrium noise level strategy is in the interior of the set $\Sigma_{\mathbf{1}, \mathbf{1}}$. Given that the noise level strategy is in the interior of $\Sigma_{\mathbf{1}, \mathbf{1}}$, there exists $i \in [K]$ such that by decreasing its noise level, we remain in the set $\Sigma_{\mathbf{1}, \mathbf{1}}$ (i.e., the user still finds it optimal to share with all platforms), but this deviation increases the payment and therefore the utility of platform $i$. Also, \cref{Lem:submodular} guarantees that the user still finds it optimal to share with all platforms.

To gain the intuition of the second part of \cref{thm:equilibirum:K>2}, notice that for small enough $\beta$, the data buyer's gain when the user's data is revealed to them is small and therefore, the buyer is willing to pay only a low price for the user data. This means that high-cost platforms will never find it optimal to enter the market. Nonetheless, the low-cost platforms always enter the market. The fact that the user shares with all the low-cost platforms (that have entered), the buyer purchases from them, and the user utility becomes zero in equilibrium follows from a similar argument to the one discussed above. 

{
\begin{remark}
While \Cref{thm:equilibirum:K>2} focuses on the equilibrium characterization for small and sufficiently large values of $\beta$, in \Cref{section:extension_intermidiary} we further characterize the equilibrium for \textit{intermediate} values of $\beta$. In particular, we establish the existence of a sequence of equilibria in which the low-cost platforms always enter the market, and as $\beta$ increases, the high-cost platforms enter sequentially, in the order of their costs (see \Cref{theorem:intermidiary_exists}). Furthermore, we show that for each high-cost platform, there exists a minimum value of $\beta$ required for market entry, and this threshold increases linearly with the platform’s cost of production (see \Cref{proposition:intermidiary_LB}).
 
\end{remark}
}

\cref{thm:equilibirum:K>2} leads to the following insights. 






\paragraph*{Competition among platforms helps the buyer and not the user:} Let us first characterize the equilibrium when we have only one platform. 

\begin{proposition}\label{Pro:Equilibirum:K=1}
Suppose Assumption \ref{assumption:substitute} holds. When $K=1$, we have the following: 
    \begin{itemize}
        \item Suppose $\alpha \le 1/\gamma_1^2$. If $c_1 \le 1/2+\beta \gamma_1^2 /2$ (which holds whenever $c_1\le 1/2$), then in equilibrium, the platform enters the market and chooses $\sigma_1=0$, the user shares with the platform, and the buyer purchases the data. In this case, the user utility is $(1-\alpha \gamma_1^2)/2$, and the buyer utility is zero. Otherwise, if $c_1 > 1/2+\beta \gamma_1^2/2$, then the platform does not enter the market, and all utilities are zero. 
        \item Suppose $\alpha > 1/\gamma_1^2$. If $c_1 \le 1/2 + \beta/2 \alpha$, then in equilibrium, the platform enters the market and chooses $\sigma_1=\sqrt{2(\alpha \gamma_1^2-1)}$, the user shares with the platform and the buyer purchases the data. In this case, both the user utility and the buyer utility are zero. Otherwise, if $c_1 > 1/2 + \beta/2 \alpha$, then the platform does not enter the market, and all utilities are zero.
    \end{itemize}   
\end{proposition}
We are interested in the second regime where $\alpha$ is large enough so that the user is not incentivized to share with the platform, irrespective of the noise level choice of the platform. In this case, as established in \cref{Pro:Equilibirum:K=1}, both the user and the data buyer utilities are zero. Now, let us compare this setting to a setting with $K \ge 2$ platforms that are competing to obtain the user's data and sell it to the data buyer, established in \cref{thm:equilibirum:K>2}. Interestingly, we observe that competition does not help the user (as the user obtains zero utility for any $K$). However, increasing the number of platforms from $K=1$ to $K\ge 2$ improves the data buyer's utility. To gain the intuition for this observation, notice that the user's utility will always be zero because otherwise, the platforms can always decrease their noise levels by a small margin so that the user still shares and increases their payments. Now, why does the competition help the data buyer? This is because of the data externality among the platforms' data about the user, similar to that of \cite{acemoglu2019too} and \cite{bergemann2020economics}. In particular, the data of a platform (because of the submodularity of the revealed information, proved in Lemma \ref{Lem:submodular}) decreases the marginal value of another platform's data, and therefore, the platforms end up competing with each other, and sell their data at lower marginal prices.


\paragraph*{More platforms imply higher utilitarian welfare:} Let us first find the overall utilitarian welfare in equilibrium, where utilitarian welfare is defined as the sum of the utilities of the platforms, the user, and the data buyer.  
\begin{corollary}\label{Cor:welfare}
   Consider $\bar{\alpha}$, $\bar{\beta}$, and $\underline{\beta}$ established in \cref{thm:equilibirum:K>2}. For $\alpha \ge \bar{\alpha}$, we have the following:
\begin{itemize}
    \item If $\beta \ge \bar{\beta}$, the utilitarian welfare in equilibrium becomes 
    \begin{align*}
        K \left(\frac{1}{2} + \frac{\beta}{ 2 \alpha} \right) - \sum_{i=1}^K c_i.
    \end{align*}
    \item If $\beta \le \underline{\beta}$, the utilitarian welfare in equilibrium becomes 
    \begin{align*}
        (\text{number of low-cost platforms})\left(\frac{1}{2} + \frac{\beta}{ 2 \alpha} \right) - \sum_{i \in [K]~:~c_i< \frac{1}{2}} c_i.
    \end{align*}
\end{itemize}  
\end{corollary}
Notably, \cref{Cor:welfare} shows that, within the ranges of each case, the overall utilitarian welfare is increasing in $\beta$ and decreasing in $\alpha$. This is because the payments always cancel each other out, and as $\beta$ increases, the data buyer's gain becomes larger, and as $\alpha$ increases, the user's loss becomes larger. Moreover, in the first case, the utilitarian welfare is increasing in the number of platforms, and in the second case, it is increasing in the number of low-cost platforms. 

\section{Regulation}\label{sec:regulations}
In this section, we consider regulation of the three-layer data market to improve the user utility. As we have established in \cref{thm:equilibirum:K>2}, the user utility in equilibrium is zero. We now explore whether we can improve the user utility by imposing some form of regulation. As discussed in the introduction, one possible regulation is to impose a limit on the amount of information each platform can leak about the user to the data buyer. This means imposing a lower bound on the noise level. Let us formally define this regulation.
\begin{definition}[Minimum privacy mandate]\label{def:regulation:lowerbound}
\sloppy   A minimum privacy mandate, represented by $\bar{\bsigma}=(\bar{\sigma}_1, \dots, \bar{\sigma}_K)$, is a regulation that mandates that each platform $ i \in [K]$ must choose a noise level above $\bar{\sigma}_i \in \mathbb{R}_+$. A special case of this regulation is uniform minimum privacy mandate is such that $\bar{\sigma}_i=\bar{\sigma}$ for all $i \in [K]$. Another special case of this regulation is to ban the platforms from sharing with the buyer, i.e.,  $\bar{\sigma}_i=\infty$ for all $i \in [K]$. 
\end{definition}
\paragraph*{Uniform minimum privacy mandate versus ban:} One may conjecture that banning the platforms from sharing their data with the buyer is user-optimal in the class of all uniform minimum privacy mandate regulations because it reduces the privacy loss of the user to zero. We next prove that this is not necessarily true. The main intuition is that if we ban data sharing, then only low-cost platforms enter the market. This, in turn, implies that the user does not obtain the gain from the services provided by the high-cost platforms, and this loss in the service gain can dominate the gain in privacy loss. We next formalize this intuition. 

\begin{proposition}\label{pro:uniform:mandate:versus:ban}
     Consider $\alpha \ge \bar{\alpha}$ for $\bar{\alpha}$ established in \cref{thm:equilibirum:K>2}. We have:
     \begin{itemize}
         \item If all platforms are low-cost, then irrespective of the minimum privacy mandate, all platforms enter the market in equilibrium. In this case, banning data sharing achieves a higher user utility than any (uniform or non-uniform) minimum privacy mandate.
         \item Otherwise, if there is at least one high-cost platform, then there exist $\tilde{\beta}$, $\underline{\beta}$, and $\bar{\sigma}$ such that
         \begin{itemize}
             \item If $\beta \ge \tilde{\beta}$, then under uniform minimum privacy mandate $\bar{\sigma}$ all platforms enter the market, and the user utility is higher than banning data sharing. 
             \item If $\beta \le \underline{\beta}$, then irrespective of the minimum privacy mandate, only low-cost platforms enter the market. In this case, banning data sharing achieves a higher user utility than any uniform minimum privacy mandate. 
        \end{itemize}
     \end{itemize}
\end{proposition}
The first part is rather straightforward: if all platforms are low-cost, then they all enter the market and provide services to the user. Banning data sharing is user-optimal because it decreases the privacy loss of the user to zero while the user obtains the service gains from all platforms. The second part is more nuanced: here, if $\beta$ is large enough, then all platforms enter the market. Now, there are two opposing forces in place. If we ban data sharing, only low-cost platforms enter the market, and user utility then comprises only the service gained from low-cost platforms, while the privacy loss becomes zero. Therefore, banning data sharing decreases the service gains but also decreases the privacy losses. If we do not ban data sharing, the user utility comprises the service gains from all platforms and the privacy loss incurred by the data sharing of all platforms with the buyer. Therefore, not banning data sharing increases the service gains but also increases the privacy losses. For a large enough minimum uniform privacy mandate, and when there is at least one high-cost platform, the service gains of the user utility dominate the privacy loss, and therefore, not banning data sharing becomes the user-optimal regulation. Finally, to understand the last part, we observe that for small enough $\beta$, only low-cost platforms enter the market, and therefore, the situation is effectively the same as the first part of \cref{pro:uniform:mandate:versus:ban}. Notice that the assumption of large enough $\alpha$ and large/small enough $\beta$ is adopted for the same reason as that of \cref{thm:equilibirum:K>2} so that an equilibrium exists.

\paragraph*{Non-uniform versus uniform minimum privacy mandate:} As noted above, when all platforms are low-cost or when $\beta$ is small enough so that high-cost platforms do not enter the market, then banning data sharing is optimal in the class of all (uniform or non-uniform) minimum privacy mandates.
However, corresponding to the second part of \cref{pro:uniform:mandate:versus:ban}, the user prefers a minimum privacy mandate to banning data sharing when there is at least one high-cost platform, and $\beta$ is large enough. This is because if we ban data sharing, the high-cost platforms do not enter, while with the ``right'' choice of uniform minimum privacy mandate, they enter the market. In this case, the user benefits from the services, while the privacy loss that they incur is small. However, with a uniform minimum privacy mandate, the low-cost platforms (that enter the market regardless of the minimum privacy mandate in place) also bring about privacy loss to the user. Therefore, a natural idea is to have a non-uniform minimum privacy mandate to help the user further. We next formalize this idea.

\begin{proposition}\label{pro:optimal:mandate}
     Suppose there is at least one high-cost platform and consider $\alpha \ge \bar{\alpha}$ for $\bar{\alpha}$ established in \cref{thm:equilibirum:K>2}. 
     There exists $\hat{\beta}$ and $\bar{\sigma}$ such that for $\beta \ge \hat{\beta}$ banning the low-cost platforms from data sharing and imposing minimum privacy mandate equal to $\bar{\sigma}$ for all high-cost platforms generates a higher user utility than any non-uniform minimum privacy mandate. 
\end{proposition}
With the non-uniform minimum privacy mandate described above, the user obtains the service gain of the low-cost platforms and incurs zero privacy loss from them. Moreover, for large enough $\beta$, the high-cost platforms all enter the market. Therefore, the user gains from the high-cost platforms' services, and when $\bar{\sigma}$ is large enough, it incurs a small privacy loss. Notice that, as established in \cref{pro:uniform:mandate:versus:ban}, with the optimal uniform minimum privacy mandate, both the low-cost and high-cost platforms enter the market. We prove that if we impose the same minimum privacy mandate but only for high-cost platforms and ban the low-cost platforms, the user obtains the same service gain from all platforms but incurs a smaller privacy loss. Moreover, all platforms still find it optimal to enter the market.

\cref{pro:uniform:mandate:versus:ban} and \cref{pro:optimal:mandate} establish that banning the data market is not necessarily user-optimal, especially when we have high-cost platforms (i.e., smaller platforms whose cost of providing the service is high). Moreover, a uniform minimum privacy mandate improves the user utility because it also incentivizes the high-cost platforms to enter the market to provide service to the user. Finally, a non-uniform minimum privacy mandate that bans the low-cost (i.e., large) platforms from data sharing and imposes a minimum privacy mandate on high-cost (i.e., small) platforms further improves the user utility. This is because this regulation not only incentivizes the high-cost platforms to enter the market to provide service to the user but also zeros out the privacy loss of the low-cost platforms.


\section{Conclusion}\label{sec:conclusion}

We have studied the intricate interplay between user privacy concerns and the dual incentives of online platforms: to safeguard user privacy and to monetize user data through sales to third-party buyers. To this end, we have constructed a multi-stage game model that captures the interactions between users, platforms, and data buyers. Our model delineates various scenarios where the interplay of platform service costs, the valuation of data, and privacy concerns dictate equilibrium outcomes. Additionally, we examine potential regulations designed to bolster user privacy and enhance market outcomes for users.

There are several exciting avenues for future research. In our framework, we have modeled a user as a representative of a market segment. Extending our analysis to a setting with multiple heterogeneous users would be very worthwhile. This is because data externalities among different users may cause indirect information disclosure, which in turn could alter both user strategies and potential regulations aimed at enhancing user experience. Similarly, our focus has been on a representative data buyer who aims to acquire user data and learn their preferences. It would be useful to consider an elaborated model with multiple buyers, given that user data is inherently multi-dimensional and can be aggregated from diverse sources, potentially modifying user strategies and calling for tailored regulation. Lastly, we have investigated regulation that restricts the transfer of user data to buyers. The challenge of developing alternative regulatory frameworks that address the intricacies of the data market—frameworks that are both adaptable and resilient enough to keep pace with swift technological advancements—will remain of significant importance for quite some time to come.

\section{Acknowledgment}
Alireza Fallah acknowledges support from the European Research Council Synergy Program, the National Science Foundation under grant number DMS-1928930, and the Alfred P. Sloan Foundation under grant G-2021-16778. The latter two grants correspond to his residency at the Simons Laufer Mathematical Sciences Institute (formerly known as MSRI) in Berkeley, California, during the Fall 2023 semester. Michael Jordan acknowledges support from the Mathematical Data Science program of the Office of Naval Research under grant number N00014-21-1-2840 and the European Research Council Synergy Program.


\bibliography{References}  


\newpage
\appendix
\section{Missing Proofs}\label{App:A}

This appendix includes the omitted proofs from the text. 

\subsection*{Proof of \cref{Lem:revealed:closedform}}
{Without loss of generality, we prove the claim for $S = [K]$.  
We first recall a property of the multivariate normal distribution:  
If $\mathbf{x}$ is an $n$-dimensional normal random variable partitioned as
\[
\mathbf{x} =
\begin{bmatrix}
\mathbf{x}_1 \\[2pt]
\mathbf{x}_2
\end{bmatrix},
\]
with corresponding mean and covariance matrices
\[
\boldsymbol\mu =
\begin{bmatrix}
\boldsymbol\mu_1 \\[2pt]
\boldsymbol\mu_2
\end{bmatrix},
\quad
\boldsymbol\Sigma =
\begin{bmatrix}
\boldsymbol\Sigma_{11} & \boldsymbol\Sigma_{12} \\[2pt]
\boldsymbol\Sigma_{21} & \boldsymbol\Sigma_{22}
\end{bmatrix},
\]
then the conditional distribution of $\mathbf{x}_1 \mid \mathbf{x}_2 = \mathbf{a}$ is normal with covariance matrix
\[
\boldsymbol\Sigma_{11} - \boldsymbol\Sigma_{12}\boldsymbol\Sigma_{22}^{-1}\boldsymbol\Sigma_{21}.
\]

\smallskip
Applying this result to our setting, let $x_1 = \mathbf{y}^\top \boldsymbol\theta$ and 
$\mathbf{x}_2 = (\tilde{\mathbf{S}}_i : i \in S)$, where each 
$\tilde{\mathbf{S}}_i = \mathbf{x}_i^\top \boldsymbol\theta + \mathcal{N}(0,1) + \mathcal{N}(0,\sigma_i^2)$.  
Then:
\begin{itemize}
    \item $\Sigma_{11}$ is the variance of $\mathbf{y}^\top \boldsymbol\theta$, which equals $1$.
    \item $\Sigma_{22}$ is a $|S|\times|S|$ matrix with diagonal entries 
    $\mathrm{Var}(\tilde{\mathbf{S}}_i) = 2 + \sigma_i^2$, 
    and off-diagonal entries 
    $\mathrm{Cov}(\tilde{\mathbf{S}}_i, \tilde{\mathbf{S}}_j) = \mathbf{x}_i^\top \mathbf{x}_j = \gamma_i \gamma_j$ 
    by Assumption~\ref{assumption:substitute}.
    \item $\Sigma_{12}$ is a $1\times|S|$ vector whose $i$-th entry is 
    $\mathrm{Cov}(\mathbf{y}^\top \boldsymbol\theta,\tilde{\mathbf{S}}_i) = \mathbf{y}^\top \mathbf{x}_i = \gamma_i$.
\end{itemize}

Therefore, the posterior variance of $Z = \mathbf{y}^\top \boldsymbol\theta$ conditional on $(\tilde{\mathbf{S}}_i : i\in S)$ is
\[
\mathrm{Var}(Z \mid (\tilde{\mathbf{S}}_i : i\in S))
= 1 - \mathbf{m}^\top M^{-1}\mathbf{m},
\]
where $\mathbf{m} = (\gamma_i)_{i\in S}$ and $M$ is the $|S|\times|S|$ matrix with
$[M]_{ii} = 2 + \sigma_i^2$ and $[M]_{ij} = \gamma_i\gamma_j$ for $i\neq j$.  

Thus, by definition, the revealed information becomes 
\begin{align*}
\mathcal{I}(\boldsymbol{\sigma}_S)&=\mathbb{E}\left[ \left( Z -\mathbb{E} _{Z \sim \pi}\left[ Z \right] \right)^{2}\right]- \mathbb{E} \left[ \left( Z -\mathbb{E}_{Z \sim \pi _{\mathcal{S}}} \left[ Z \right] \right) ^{2}\right] \\
&= \mathrm{Var}(Z)- \mathrm{Var}(Z \mid (\tilde{\mathbf{S}}_i : i\in S))\\
& = 1- \left(1 -\mathbf{m}^\top M^{-1}\mathbf{m} \right)=\mathbf{m}^\top M^{-1}\mathbf{m},
\end{align*}
which completes the proof. $\blacksquare$

}

\subsection*{Proof \cref{Lem:monotone}}
{We make use of Lemma \ref{Lem:revealed:closedform} in this proof. Throughout, $M$ is a covariance matrix defined in that lemma.

\noindent \textbf{Monotonicity in the set $S$.}
 Let $T=S\cup\{k\}$ where $k\notin S$. We use the subscripts $S$ and $T$ to indicate the dependence on the set and write the block decomposition.
\[
M_T=\begin{bmatrix}
M_S &&& \bm u \\[2pt]
\bm u^\top &&& a
\end{bmatrix},\qquad
\bm m_T=\begin{bmatrix}\bm m_S\\ m_k\end{bmatrix},
\]
where $\bm u=(\gamma_i\gamma_k)_{i\in S}$ and $a=2+\sigma_k^2$. Since $M_S\succ0$, the Schur complement
\[
\Delta \;\equiv\; a-\bm u^\top M_S^{-1}\bm u \;>\; 0.
\]
The block matrix inversion formula yields
\[
M_T^{-1}=
\begin{bmatrix}
M_S^{-1}+M_S^{-1}\bm u\,\Delta^{-1}\bm u^\top M_S^{-1} & -M_S^{-1}\bm u\,\Delta^{-1}\\[2pt]
-\Delta^{-1}\bm u^\top M_S^{-1} & \Delta^{-1}
\end{bmatrix}.
\]
Therefore, we can write 
\begin{align*}
\mathcal{I}(\bsigma_T)=\bm m_T^\top M_T^{-1}\bm m_T=\bm m_S^\top M_S^{-1}\bm m_S
+ \big(m_k-\bm u^\top M_S^{-1}\bm m_S\big)^2\,\Delta^{-1}.
\end{align*}
Since $\Delta^{-1}>0$, the second term is nonnegative, implying
\[
\mathcal{I}(\bsigma_T)\;-\;\mathcal{I}(\bsigma_S)
=\frac{\big(m_k-\bm u^\top M_S^{-1}\bm m_S\big)^2}{\,a-\bm u^\top M_S^{-1}\bm u\,}\;\ge 0.
\]
Therefore, $\mathcal{I}(\bsigma_S)$ is increasing in $S$.

\noindent \textbf{Monotonicity in $\bsigma_S$.} 
Let us denote $\mathcal{I}(\boldsymbol{\sigma}_S)=\mathbf{m}^T M^{-1}\mathbf{m}$ with $M=M(\boldsymbol{\sigma}_S)$ depending on $\sigma_i$ only through the diagonal entries. Therefore, we have $\partial M/\partial \sigma_i=2\sigma_i \bm e_i \bm e_i^\top$, where $\bm e_i$ is the $i$th standard basis vector (for $i\in S$). Using the matrix derivative identity
\[
\frac{\partial M^{-1}}{\partial \sigma_i}
=-M^{-1}\!\left(\frac{\partial M}{\partial \sigma_i}\right)\!M^{-1},
\]
we obtain
\begin{align*}
\frac{\partial \mathcal{I}(\boldsymbol{\sigma}_S)}{\partial \sigma_i}
&=\bm m^\top\!\left(\frac{\partial M^{-1}}{\partial \sigma_i}\right)\!\bm m
=-\bm m^\top M^{-1}\!\left(2\sigma_i\,\bm e_i\bm e_i^\top\right)\!M^{-1}\bm m\\
&=-2\sigma_i\,\big(\bm e_i^\top M^{-1}\bm m\big)^2\;\le\;0.
\end{align*}
Therefore, $\mathcal{I}(\bsigma_S)$ is decreasing in each $\sigma_i$. $\blacksquare$ 
}
\subsection*{Proof of \cref{Lem:submodular}}
Using Lemma \ref{Lem:revealed:closedform}, we first derive the leaked information to the buyer. In particular, if both platforms enter the market and the user's data is shared with them and then sold to the buyer, then the leaked information is given by 
\begin{align*}
\mathcal{I}(\sigma_1, \sigma_2) = \frac{\gamma_1^2(2+\sigma_2^2) + \gamma_2^2 (2+\sigma_1^2)- 
   2\gamma_1^2 \gamma_2^2}{(2+\sigma_1^2) (2+ \sigma_2^2) - \gamma_1^2 \gamma_2^2}.
\end{align*}
On the other hand, if the user's data is only shared with the buyer through platform $i$, then the leaked information to the buyer is given by  
\begin{align*}
    \mathcal{I}(\sigma_i) = \frac{\gamma_i^2}{2+\sigma_i^2} .
\end{align*}
To prove this lemma, without loss of generality, we need to establish that if the buyer has the extra data of platform $j$, the gain in acquiring the data of platform $i$ is smaller. By invoking the fact mentioned in the proof of Lemma \ref{Lem:revealed:closedform} for the normal variable $(\by^T \btheta, \bx_i^T \btheta + \mathcal{N}(0,1)+\mathcal{N}(0, \sigma_i^2), \bx_j^T \btheta + \mathcal{N}(0,1)+\mathcal{N}(0, \sigma_j^2))$ conditional on the rest of the data points, the inequality stated in the lemma becomes equivalent to 
\begin{align*}
    \frac{\tilde{\gamma}_i^2}{2+\sigma_i^2} \ge \frac{\tilde{\gamma}_i^2(2+\sigma_j^2)+ \tilde{\gamma}_j^2(2+\sigma_i^2)-2  \tilde{\gamma}^2_i \tilde{\gamma}^2_j }{(2+\sigma_i^2) (2+\sigma_j^2)- \tilde{\gamma}_{i}^2\tilde{\gamma}_j^2} - \frac{\tilde{\gamma}_j^2}{2+\sigma_j^2},
\end{align*}
where for all $\ell \in [K]$, $\tilde{\gamma}_{\ell}=\gamma_{\ell} (\tilde{\mathrm{var}})^{1/2}$ where $\tilde{\mathrm{var}}$ is the variance of $\btheta^T \btheta$ conditional on the rest of the data points. The above inequality holds true as it is equivalent to
\begin{align*}
\tilde{\gamma}_{i}^2\tilde{\gamma}_j^2 \left(\frac{\tilde{\gamma}_i^2}{2+\sigma_i^2} + \frac{\tilde{\gamma}_j^2}{2+\sigma_i^2} \right) \le 2 \tilde{\gamma}_{i}^2\tilde{\gamma}_j^2
\end{align*}
that holds as $\tilde{\gamma}_i, \tilde{\gamma}_j \le 1$. $\blacksquare$

\subsection*{Proof of \cref{Lem:submodular:sigma}}
We prove part one and part two follows an identical argument. We use a similar argument to the one presented in the proof of \cref{Lem:submodular}. We establish this inequality by considering a one-by-one change in the noise variance. 


If $\bsigma_{-j}=\bsigma'_{-j}$ and ${\sigma'_j}^2\ge \sigma_j^2$, then the inequality becomes equivalent to 
\begin{align*}
  &\frac{\tilde{\gamma}_i^2(2+\sigma_j^2)+ \tilde{\gamma}_j^2(2+\sigma_i^2)-2  \tilde{\gamma}^2_i \tilde{\gamma}^2_j }{(2+\sigma_i^2) (2+\sigma_j^2)- \tilde{\gamma}_{i}^2\tilde{\gamma}_j^2} - \frac{\tilde{\gamma}_j^2}{2+\sigma_j^2} \\
  & \le \frac{\tilde{\gamma}_i^2(2+{\sigma'_j}^2)+ \tilde{\gamma}_j^2(2+\sigma_i^2)-2  \tilde{\gamma}^2_i \tilde{\gamma}^2_j }{(2+\sigma_i^2) (2+{\sigma'_j}^2)- \tilde{\gamma}_{i}^2\tilde{\gamma}_j^2} - \frac{\tilde{\gamma}_j^2}{2+{\sigma'_j}^2}.
\end{align*}
The above inequality holds because the function 
\begin{align*}
    x \mapsto \frac{\tilde{\gamma}_i^2(2+x)+ \tilde{\gamma}_j^2(2+\sigma_i^2)-2  \tilde{\gamma}^2_i \tilde{\gamma}^2_j }{(2+\sigma_i^2) (2+x)- \tilde{\gamma}_{i}^2\tilde{\gamma}_j^2} - \frac{\tilde{\gamma}_j^2}{2+x}
\end{align*}
is increasing, which can be seen by evaluating its derivative. $\blacksquare$

\subsection*{Proof of \cref{Pro:Equilibirum:Buyer:Price}}

We first argue that, for any $i \in [K]$ with $a_i=e_i=1$, the equilibrium price of platform $i$ must be such that the buyer finds it optimal to purchase from this platform in the corresponding buyer equilibrium. To see this notice that if the buyer does not purchase from platform $i$, the platform's utility from interacting with the buyer becomes $0$. Now, if the platform deviates and selects the price
\begin{align*}
    p_i=\beta \left( \mathcal{I}(\bsigma, \be, \ba, \mathbf{1}) - \mathcal{I}(\bsigma, \be, \ba, (b_i=0,\mathbf{1}_{-i})) \right)
\end{align*}
then the buyer always finds it optimal to purchase from this platform, and therefore, the platform's utility weakly increases from $0$. This is because for any $\bb_{-i}$, we have
\begin{align*}
&\mathcal{U}_{\text{buyer}}(\bsigma, \be, \ba, (p_i,\bp_{-i}), (b_i=1,\bb_{-i}))- \mathcal{U}_{\text{buyer}}(\bsigma, \be, \ba, (p_i,\bp_{-i}), (b_i=0,\bb_{-i})) \\
&= \beta \left[\left( \mathcal{I}(\bsigma, \be, \ba, (b_i=1,\bb_{-i})) - \mathcal{I}(\bsigma, \be, \ba, (b_i=0,\bb_{-i})) \right) \right. \\
& \quad \left. 
-  \left( \mathcal{I}(\bsigma, \be, \ba, \mathbf{1}) - \mathcal{I}(\bsigma, \be, \ba, (b_i=0,\mathbf{1}_{-i})) \right) \right] \overset{(a)}{\ge}0
\end{align*}
where (a) follows from \cref{Lem:submodular}. 
{ This establishes that, at equilibrium, the buyer will purchase data from all platforms that have entered the market and possess user data.

Now, suppose $\bp$ is a price equilibrium. Fix $i$. On the one hand, the buyer has no incentive to deviate and refrain from buying data from the platform $i$ if and only if
\begin{equation*} 
p_i \leq  \beta \left( \mathcal{I}(\bsigma, \be, \ba, \mathbf{1}) - \mathcal{I}(\bsigma, \be, \ba, (b_i=0,\mathbf{1}{-i})) \right).
\end{equation*}
On the other hand, if this inequality is not strict, then the platform $i$ would have an incentive to increase its price. However, if the price is equal to
\begin{equation*}
\beta \left( \mathcal{I}(\bsigma, \be, \ba, \mathbf{1}) - \mathcal{I}(\bsigma, \be, \ba, (b_i=0,\mathbf{1}{-i})) \right),
\end{equation*}
platform also has no incentive to deviate. This implies that the prices given in \Cref{Pro:Equilibirum:Buyer:Price} constitute an equilibrium and that this is the unique price equilibrium. This completes the proof.
}
$\blacksquare$
\subsection*{Proof of \cref{Lem:equilibrium:lattice}}
This lemma directly follows the submodularity of the leaked information in actions, established in \cref{Lem:submodular}. $\blacksquare$
\subsection*{Proof of \cref{Pro:Equilibirum:User:Noise}}
Suppose the contrary that there exists a platform equilibrium strategy $(\bsigma, \be)$ for which $e_i=1$, $a_i=0$ and $\ba \in \mathcal{A}^{\mathrm{E}}(\bsigma, \be)$. We argue that platform $i$ has a profitable deviation. In particular, notice that the current utility of platform $i$ is $-c_i$, as the buyer will not pay any non-zero price for this platform's data. Now, if platform $i$ chooses $\tilde{e}_i=1$, $\tilde{p}_i=0$ and $\tilde{\sigma}_i=\infty$, then in any $\ba \in \mathcal{A}^{\mathrm{E}}((\tilde{\sigma}_i, \bsigma_{-i}), \be)$ we must have that $a_i=1$ and therefore the platform $i$'s utility increases to $\mathcal{I}_i-c_i$. $\blacksquare$
\subsection*{Proof of \cref{lemma:alpha_1.88}}
Recall that we argued that the equilibrium, if exists, belong to $\tsyy \cap \tsnn$. Furthermore, we stated that we only need to check deviations in Figure \ref{fig:RegionsK=2}. To make sure such a deviation does not happen for some equilibrium $(\sigma_1, \sigma_2) \in \tsyy \cap \tsnn$, we must have
\begin{align} \label{eqn:no_deviation}
    & \mathcal{U}_1((\sigma_1, \sigma_2)), \mathbf{1}, \mathbf{p}^*, \mathbf{1}) = \nonumber \\
    &\quad \frac{1}{2} + \beta \left(\mathcal{I}(\sigma_1, \sigma_2)- \mathcal{I}(\sigma_2) \right) \ge \frac{1}{2} + \beta \mathcal{I}(\tilde{\sigma}_1) = \mathcal{U}_1((\tilde{\sigma}_1, \sigma_2)), (1,0)), \mathbf{p}^*, \mathbf{1}). 
\end{align}
Notice that $(\sigma_1, \sigma_2)$ is on the boundary of $\tsyy$ and $\tsnn$ which implies that  
\begin{equation} \label{eqn:boundary_1_alpha}
\mathcal{U}_{\text{user}}(\sigma_1, \sigma_2) = 1 - \alpha \mathcal{I}(\sigma_1, \sigma_2) = 0,   
\end{equation}
and thus, $\mathcal{I}(\sigma_1, \sigma_2) = 1/\alpha$. Using this, we can rewrite the condition \eqref{eqn:no_deviation} as
\begin{equation}\label{eqn:no_deviation_2}
\frac{1}{\alpha} \geq \mathcal{I}(\sigma_2) + \mathcal{I}(\tilde{\sigma}_1).
\end{equation}
Notice that we can similarly consider the deviation of the second platform by increasing its noise variance to $\tilde{\sigma}_2^2$ in Figure \ref{fig:RegionsK=2}. To ensure this deviation is also not profitable, we should have
\begin{equation}\label{eqn:no_deviation_3}
\frac{1}{\alpha} \geq \mathcal{I}(\sigma_1) + \mathcal{I}(\tilde{\sigma}_2).
\end{equation}
The following lemma establishes the necessary and sufficient condition for \eqref{eqn:no_deviation_2} and \eqref{eqn:no_deviation_3} to hold for some $(\sigma_1, \sigma_2) \in \tsyy \cap \tsnn$.

Notice that if we move up on the boundary $\tsyy \cap \tsnn$, $\sigma_2$ and $\tilde{\sigma}_1$ both would increase, and hence,
$\mathcal{I}(\sigma_2)$ and $ \mathcal{I}(\tilde{\sigma}_1)$ both would decrease. Therefore, if \eqref{eqn:no_deviation_2} holds for some $(\sigma_1, \sigma_2) \in \tsyy \cap \tsnn$, then it would hold for any upper point on the boundary. Similarly, if \eqref{eqn:no_deviation_3} holds for some $(\sigma_1, \sigma_2) \in \tsyy \cap \tsnn$, then it would hold for any lower point on the boundary. Hence, if a point satisfies both conditions, then for any other point on the boundary $\tsyy \cap \tsnn$, at least one of the conditions would hold.   

Next, we show that, for the point given by \eqref{eqn:equilibrium_K=2}, both conditions hold if $\alpha \geq \bar{\alpha}$ and neither hold if $\alpha < \bar{\alpha}$. Given our discussion above, this would show that for $\alpha < \bar{\alpha}$, there is no point for which both conditions hold and hence the proof would be complete. 

\begin{figure}[t]
    \centering
    \includegraphics[scale=.5]{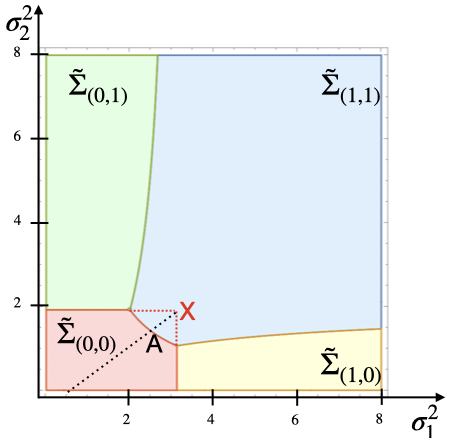}
    \caption{Illustrating the point given by \eqref{eqn:equilibrium_K=2}.}
    \label{fig:good_point}
\end{figure}

Now, let us first see why the point given by \eqref{eqn:equilibrium_K=2} is actually on the boundary $\tsyy \cap \tsnn$. Consider point $X$ in Figure \ref{fig:good_point}. For this point, we have 
\begin{equation*}
\frac{\gamma_1^2}{2+\sigma_1^2} = \frac{\gamma_2^2}{2+\sigma_2^2} = \frac{1}{2\alpha}.  
\end{equation*}
Hence, the line 
\begin{equation} \label{eqn:line_symmetric_K=2}
\frac{2+\sigma_1^2}{\gamma_1^2} = \frac{2+\sigma_2^2}{\gamma_2^2}    
\end{equation}
passes through point $X$, and therefore, it intersects with the boundary $\tsyy \cap \tsnn$. Let us call this intersection $A$ (see Figure \ref{fig:good_point}) and denote it by $(\sigma_1^A, \sigma_2^A)$. Using \eqref{eqn:boundary_1_alpha} along with the fact that $A$ is on the line \eqref{eqn:line_symmetric_K=2}, we can derive that
\begin{equation} \label{eqn:equi_point}
\frac{2+(\sigma_1^A)^2}{\gamma_1^2} = \frac{2+(\sigma_2^A)^2}{\gamma_2^2}  = 2\alpha-1.
\end{equation}
Now, let us consider the deviation to $(\tilde{\sigma_1}, \sigma_2^A)$ which is on the boundary of $\tsyy$ and $\tsyn$, i.e.,
\begin{equation} \label{eqn:deviation_point}
1 - \alpha  \mathcal{I}(\tilde{\sigma_1}, \sigma_2^A)  = \frac{1}{2} - \alpha \mathcal{I}(\tilde{\sigma_1}).   
\end{equation}
Let 
\begin{equation} \label{eqn:deviation_ratio}
\eta = \frac{\gamma_1^2}{2+\tilde{\sigma_1}^2}.    
\end{equation}
Plugging \eqref{eqn:equi_point} and \eqref{eqn:deviation_ratio} into \eqref{eqn:deviation_point} and simplifying the equation yields
\begin{equation}
\eta \in \left\{ 
1 - \frac{1}{4\alpha} - \frac{\sqrt{16 \alpha^2 - 16\alpha +1}}{4\alpha},
1 - \frac{1}{4\alpha} + \frac{\sqrt{16 \alpha^2 - 16\alpha +1}}{4\alpha}
\right\}.    
\end{equation}
Next we argue that we should pick the smaller (first) solution. Notice that since $\tilde{\sigma}_1 \geq \sigma_1^A$, we have $\eta = \mathcal{I}(\tilde{\sigma}_1) \leq \mathcal{I}(\sigma_1^A) = 1/(2\alpha-1)$. In addition, Assumption \ref{assumption:all_regions}, or equivalently \eqref{eqn:assumption_K=2}, implies $\alpha \geq 3/2$. In this regime, we have $1-\frac{1}{4\alpha} > \frac{1}{2\alpha-1}$. Thus, we should pick the smaller solution above.

Now, we check the condition \eqref{eqn:no_deviation_2}. Notice that we need to find $\alpha$ such that 
\begin{equation} \label{eqn:condition_alpha}
\eta + \frac{1}{2\alpha-1} \leq \frac{1}{\alpha}.    
\end{equation}
We can verify that the right-hand side minus the left-hand side is equal to zero at $\bar{\alpha} \approx 1.88$ and positive afterward. Notice that $\eta$ and therefore the condition \eqref{eqn:condition_alpha} is symmetric with respect to both platforms, and consequently, considering the other deviation to $(\sigma_1^A, \tilde{\sigma}_2)$ leads to the exact same condition on $\alpha$. This completes the proof of our claim that both conditions hold if $\alpha \geq \bar{\alpha}$ and neither if $\alpha < \bar{\alpha}$. $\blacksquare$
\subsection*{Proof of \cref{eqn:proposition_equilibrium_pp}}
Suppose an equilibrium $(\sigma_1^*, \sigma_2^*)$ exists in which both platforms enter the market. As we argued in \cref{sec:Equilibrium:K=2}, in this case $(\sigma_1^*, \sigma_2^*) \in \tsyy \cap \tsnn$. Next, notice that for any $i \in \{1,2\}$, the platform $i$'s utility should be nonnegative. Therefore, we have
\begin{equation}
\frac{1}{2} - c_i + \beta (\mathcal{I}(\sigma_1^*, \sigma_2^*) - \mathcal{I}(\sigma_{-i}^*) ) \geq 0.    
\end{equation}
Notice that, since $(\sigma_1^*, \sigma_2^*) \in \tsyy \cap \tsnn$, we have $\mathcal{I}(\sigma_1^*, \sigma_2^*) = 1/\alpha$. Also, the minimum of $\mathcal{I}(\sigma_{-i}^*)$ corresponds to point $B$ for $i=1$ and point $C$ for $i=2$ in Figure \ref{fig:OneParticipate} and is equal to $1/(2\alpha)$. This implies the lower bound on $\beta$.

Now, suppose $\beta \geq \bar{\beta}$. We claim the point $(\sigma_1^A, \sigma_2^A)$, defined in the proof of \cref{lemma:alpha_1.88}, can be an equilibrium.  We just need to check when the platform's utility is nonnegative. In particular, the first platform's utility at $(\sigma_1^A, \sigma_2^A)$ is given by
\begin{equation}
\frac{1}{2} - c_1 + \beta \left (\mathcal{I}(\sigma_1^A, \sigma_2^A) - \mathcal{I}(\sigma_2^A) \right) = \frac{1}{2} + \frac{\beta(\alpha-1)}{\alpha(2\alpha-1)},
\end{equation}
which is nonnegative for  $\beta \geq \bar{\beta}$.
We can similarly write the second platform's utility at the point $A$. $\blacksquare$
\subsection*{Proof of \cref{thm:equilibirum:K>2}}
We make use of the following lemmas in this proof.

\begin{lemma}\label{Lem:interiorofSigma1}
    For any $\be$, the noise level equilibrium strategy, if it exists, belongs to the boundary of the set $\Sigma_{\mathbf{1}, \be}$. 
\end{lemma}
\emph{Proof of \cref{Lem:interiorofSigma1}:} Using \cref{Pro:Equilibirum:User:Noise}, we know that the equilibrium noise variance belongs to $\Sigma_{\mathbf{1}, \be}$. We next argue that it cannot belong to the interior of $\Sigma_{\mathbf{1}, \be}$. To see this, suppose the contrary that $\bsigma \in \mathrm{int}(\Sigma_{\mathbf{1}}, \be)$ is an equilibrium. There exists $i \in [K]$ such that, by locally decreasing $\sigma_i$, we remain in the set $\Sigma_{\mathbf{1}, \be}$. We claim that the platform $i$ has an incentive to decrease its noise variance. First, note that by decreasing its noise variance, we remain in the set $\Sigma_{\mathbf{1}, \be}$ and that the user still finds it optimal to share with platform $i$. However, the payment she receives from the buyer, which is given by 
\begin{align*}
    \beta \left( \mathcal{I}(\bsigma, \be,  \ba, \mathbf{1}) - \mathcal{I}(\bsigma, \be, \ba, (b_i=0,\mathbf{1}_{-i})) \right) 
\end{align*}
increases because of \cref{Lem:submodular:sigma}. $\blacksquare$

\begin{lemma}\label{Lem:BoundaryofSigma1}
    For any $\be$, the noise level equilibrium strategy, if it exists, belongs to $\Sigma_{\mathbf{1}, \be} \cap \Sigma_{\mathbf{0}, \be}$.
\end{lemma}
\emph{Proof of \cref{Lem:BoundaryofSigma1}:} Using \cref{Lem:interiorofSigma1}, we know that the equilibrium noise level, if it exists, belongs to the boundary of $\Sigma_{\mathbf{1}, \be}$ and $\Sigma_{\ba, \be}$ for some $\ba \in \{0,1\}^K \setminus \{\mathbf{1}\}$. We argue that it cannot belong to $\Sigma_{\mathbf{1}, \be} \cap \Sigma_{\ba, \be}$ when $\ba \neq \mathbf{0}$. Suppose the contrary that $\sigma \in \Sigma_{\mathbf{1}, \be} \cap \Sigma_{\ba, \be}$ is an equilibrium where $a_i=1$ and also $a_j=0$. We claim that platform $i$ has an incentive to decrease its noise level. In particular, we argue that by decreasing its noise level, we go to $\mathrm{int}(\Sigma_{\mathbf{1}, \be})$ and that platform $i$'s utility increases. Notice that it suffices to show that by decreasing $\sigma_i$, we remain in the set $\Sigma_{\mathbf{1}, \be}$. This is because by invoking \cref{Lem:submodular:sigma}, the payment of platform $i$ increases, and the users keep sharing with platform $i$. For $\bsigma$, the indifference condition for the user implies
\begin{align*}
    \sum_{i: ~e_i=1} a_i \mathcal{I}_i - \alpha \mathcal{I}(\bsigma, \be, \ba, \mathbf{1}) = \sum_{i: ~e_i=1} \mathcal{I}_i - \alpha \mathcal{I}(\bsigma, \be, \mathbf{1}, \mathbf{1})
\end{align*}
which can be written as
\begin{align*}
    \sum_{i:~ e_i=1, a_i=0} \mathcal{I}_i= \alpha \left(\mathcal{I}(\bsigma, \be, \mathbf{1}, \mathbf{1})- \mathcal{I}(\bsigma, \be, \ba, \mathbf{1})\right).
\end{align*}
Now if $a_i=1$, then by decreasing $\sigma_i$, the left-hand side does not change, and the right-hand decreases (from \cref{Lem:submodular:sigma}). Therefore, by decreasing $\sigma_i$, we remain in the set $\Sigma_{\mathbf{1}, \be}$. $\blacksquare$

For a given $\be$ with $e_i=1$, platform $i$ finds it optimal to enter the market if
\begin{align} \label{eqn:market_entering_condition}
    \frac{1}{2}-c_i + \beta \left(\mathcal{I}(\bsigma, \be, (a_i=1, \ba_{-i}), \mathbf{1}) - \mathcal{I}(\bsigma, \be, (a_i=0, \ba_{-i}), \mathbf{1}) \right) \ge 0.
\end{align}
Therefore, by letting 
\begin{align*}
    \beta \ge \bar{\beta}:= \sup_{i \in [K], (e_i=1, \be_{-i}), \bsigma \in \Sigma_{\mathbf{1}, \be} \cap \Sigma_{\mathbf{0}, \be}} \frac{c_i - \frac{1}{2}}{\mathcal{I}(\bsigma, \be, (a_i=1, \ba_{-i}), \mathbf{1}) - \mathcal{I}(\bsigma, \be, (a_i=0, \ba_{-i}), \mathbf{1}) },
\end{align*}
all platforms enter the market. We next argue that this term is finite. Note that, for any given value of $\be_{-i}$, the fact that $\bsigma \in \Sigma_{\mathbf{1}, \be} \cap \Sigma_{\mathbf{0}, \be}$ implies that the term 
$\mathcal{I}(\bsigma, \be, (a_i=1, \ba_{-i}), \mathbf{1})$ in the denominator is equal to 
\begin{equation} \label{eqn:revealed_info_border}
\frac{|\{i : e_i =1\}|}{2\alpha}.    
\end{equation}
As a result, we can rewrite $\bar{\beta}$ as
\begin{equation*}
\bar{\beta} = \max_{i \in [K], (e_i=1, \be_{-i})} \frac{c_i-1/2}{{|\{i : e_i =1\}|}/{(2\alpha)} - \sup_{\bsigma \in \Sigma_{\mathbf{1}, \be} \cap \Sigma_{\mathbf{0}, \be}} \mathcal{I}(\bsigma, \be, (a_i=0, \ba_{-i}), \mathbf{1}) }.    
\end{equation*}
Notice that, for any given $\be_{-i}$, the denominator is positive because the amount of leaked information strictly decreases when one less user shares information. Finally, $\bar{\beta}$ is finite becasue the outer maximum is taken over finitely many elements.


Similarly, for a given $\be$ with $e_i=1$, when $c_i>1/2$ platform $i$ finds it optimal to not enter the market if
\begin{align*}
    \frac{1}{2}-c_i + \beta \left(\mathcal{I}(\bsigma, \be, (a_i=1, \ba_{-i}), \mathbf{1}) - \mathcal{I}(\bsigma, \be, (a_i=0, \ba_{-i}), \mathbf{1}) \right) \le 0.
\end{align*}
Therefore, by letting 
\begin{align*}
    \beta \le \underline{\beta}:= \inf_{i \in [K], c_i>\frac{1}{2}, (e_i=1, \be_{-i}), \bsigma \in \Sigma_{\mathbf{1}, \be}} \frac{c_i - \frac{1}{2}}{\mathcal{I}(\bsigma, \be, (a_i=1, \ba_{-i}), \mathbf{1}) - \mathcal{I}(\bsigma, \be, (a_i=0, \ba_{-i}), \mathbf{1}) },
\end{align*}
all platforms enter the market.

We next prove that, assuming a set of platforms that have entered the market, for large enough $\alpha$, there exists an equilibrium. Without loss of generality, let us assume that all platforms have entered the market. $\bsigma \in \Sigma_{\mathbf{1},\mathbf{1} } \cap \Sigma_{\mathbf{0},\mathbf{1}}$ is an equilibrium noise level if no platform has a deviation. Let us consider platform 1. 
For this platform, the possible deviations are 
to $\Sigma_{\ba, \mathbf{1}}$ 
where $a_1=1$ and $\ba_{-1}$ can be any vector in $\{0,1\}^{K-1}$. We next argue that $\bsigma \in \Sigma_{\mathbf{1}, \mathbf{1}} \cap \Sigma_{\mathbf{0}, \mathbf{1}}$ for which 
\begin{align*}
    \frac{\gamma_i}{2+\sigma_i^2}=t \text{ for all } i\in [K]
\end{align*}
is a noise level equilibrium for large enough $\alpha$. Similar to the analysis with $K=2$ platforms, we let 
\begin{align*}
    \mathcal{I}(\sigma_{i}: i \in S)
\end{align*}
for any $S \subseteq [K]$ denote the leaked information about $\theta$ when the data of the user is shared with the buyer through platforms in the set $S$, and platform $i \in S$ is using noise variance $\sigma_i^2$.

Before proceeding with the proof, we first find a closed-form expression for the leaked information. 

\begin{lemma}\label{Lem:closed-form:g}
${}^{}$
\begin{itemize}
    \item Suppose $\frac{\gamma_i}{2+\sigma_i^2}=t$ for all $i\in [K]$. The leaked information about $\theta$ when the data of all platforms is shared with the buyer becomes
\begin{align*}
    \frac{K t}{1+ (K-1) t}.
\end{align*}
\item Suppose $\frac{\gamma_i}{2+\sigma_i^2}=t$ for all $i\in [K]\setminus \{1\}$ and $\frac{\gamma_1}{2+\sigma_1^2}=t'$. The leaked information about $\theta$ when the data of all platforms is shared with the buyer becomes
\begin{align*}
    1- \frac{1-t}{1+t'+(K-2)t}.
\end{align*}
\end{itemize}
\end{lemma}
\emph{Proof:} Using \cref{Lem:revealed:closedform}, the leaked information is 
\begin{small}
\begin{align*}
    & (\gamma_1, \dots, \gamma_K) \begin{pmatrix}
    \gamma_1^2/t     &\gamma_1 \gamma_2 \cdots &\gamma_1 \gamma_K \\
    \gamma_2 \gamma_1 & \gamma_2^2/t \cdots & \gamma_2 \gamma_K \\
     \vdots \\
     \gamma_K \gamma_1 & \gamma_K \gamma_2 \cdots & \gamma_K^2/t
    \end{pmatrix}^{-1} (\gamma_1, \dots, \gamma_K)^T\\
    \overset{(a)}{=} & \bgamma^T \left( \bgamma \bgamma^T + (\frac{1}{t}-1) \mathrm{diag}(\gamma_1^2, \dots, \gamma_K^2) \right)^{-1} \bgamma \\
    \overset{(a)}{=} &  \bgamma^T \left(\frac{1}{\frac{1}{t}-1} \mathrm{diag}((1/\gamma_i^2)_{i=1}^k)  - \frac{\frac{1}{\frac{1}{t}-1} \mathrm{diag}((1/\gamma_i^2)_{i=1}^k) ~\bgamma^T \bgamma ~ \frac{1}{\frac{1}{t}-1} \mathrm{diag}((1/\gamma_i^2)_{i=1}^k) }{1+ \bgamma^T \frac{1}{\frac{1}{t}-1}\mathrm{diag}((1/\gamma_i^2)_{i=1}^k)\bgamma } \right) \bgamma\\
    = & \frac{K t}{1+ (K-1) t},
\end{align*}
\end{small}
where in (a), we used $\bgamma=(\gamma_1, \dots, \gamma_K)^T$ and in (b), we used the Sherman–Morrison formula for the inverse of a rank-one perturbation of a matrix. This completes the proof of the first part. The second part follows from a similar argument. $\blacksquare$

Let us consider the deviation of platform 1 so that the corresponding user equilibrium is $(a_1=1, \ba_{-i})$ with $\ba_{-i}$ having $i-1$ ones and $K-i$ zeros. Notice that if such a deviation is not possible by increasing $\sigma_1$, then this is not a profitable deviation for platform 1, and we now verify that even if such a deviation is possible, it is not profitable for platform 1. Given that $\bsigma \in \Sigma_{\mathbf{1}, \mathbf{1}} \cap \Sigma_{\mathbf{0}, \mathbf{1}}$, we see that the user is indifferent between sharing with all platforms and sharing with none of them. Therefore, we have 
\begin{align*}
    \sum_{i=1}^K \mathcal{I}_i- \alpha \mathcal{I}(\sigma_1, \dots, \sigma_K)=0
\end{align*}
which gives us
\begin{align*}
    \mathcal{I}(\sigma_1, \dots, \sigma_K)= \frac{K}{ 2 \alpha}.
\end{align*}
Using \cref{Lem:closed-form:g}, the above equality gives us the following. 
\begin{align*}
    t= \frac{1}{1+ 2\alpha - K}.
\end{align*}
Given the above deviation to $(\tilde{\sigma}_1, \dots ,\sigma_K)$, the user utility gives us 
\begin{align*}
    \frac{K}{2} - \alpha \mathcal{I}(\tilde{\sigma}_1, \dots ,\sigma_K) \leq  \frac{i}{2} - \alpha \mathcal{I}(\tilde{\sigma}_1, \dots ,\sigma_i).
\end{align*}
Substituting the closed-form expressions of \cref{Lem:closed-form:g} into the above equality gives us the following.
\begin{align} \label{eqn:t'_UB}
    t'\leq \frac{2-4 \alpha + K-i + \sqrt{16 \alpha^2 +(K-i)^2-8 \alpha K}}{2+4 \alpha - 2 K},
\end{align}
where $t'=\frac{\gamma_1^2}{2+ \tilde{\sigma}_1^2}$.

Now, for this deviation to be not profitable, we must have the following.
\begin{align*}
    \mathcal{I}(\sigma_1, \dots, \sigma_K) - \mathcal{I}(\sigma_2, \dots, \sigma_K) \ge \mathcal{I}(\tilde{\sigma}_1, \dots, \sigma_i) - \mathcal{I}(\sigma_2, \dots, \sigma_i).
\end{align*}
Note that the right hand side is increasing in $t'$, and therefore, it suffices to show it holds for $t'$ given in \eqref{eqn:t'_UB}.
Again, using the closed-form expressions of \cref{Lem:closed-form:g} and writing $t'$ and $t$ in terms of $\alpha$, the above inequality becomes equivalent to
\begin{align*}
    -1 + \frac{1-K}{2 \alpha-1}+ \frac{i-1}{2 \alpha -1 + i -K} + \frac{K}{2 \alpha} + \frac{2K - 4 \alpha}{K-i + \sqrt{16 \alpha^2 + (K-i)^2 - 8 \alpha K}} \ge 0,
\end{align*}
which we prove next. We can write
\begin{align} 
   &  \frac{3K-i}{4 \alpha} + \frac{1-K}{2 \alpha -1} + \frac{i-1}{2 \alpha -1 +i -K} -1+ \frac{\sqrt{16 \alpha^2 + (K-i)^2 - 8 \alpha K}}{4 \alpha} \label{eqn:diff_two_sides_1} \\
   & \overset{(a)}{\ge}  \frac{3K-i}{4 \alpha} + \frac{1-K}{2 \alpha -1} + \frac{i-1}{2 \alpha -1 +i -K}- \frac{K}{2 \alpha \left(1+ \sqrt{1- \frac{K}{2 \alpha}} \right)} \nonumber \\
   & \ge  \frac{3K-1}{4 \alpha}- \frac{K-1}{2 \alpha }  - \frac{K}{2 \alpha \left(1+ \sqrt{1- \frac{K}{2 \alpha}} \right)}  \overset{(b)}{\ge} 0 \nonumber 
\end{align}
where (a) follows by noting that the last term is increasing in $i$ and (b) holds for
\begin{align*}
    \alpha \ge \bar{\alpha} := \frac{ (K+1)^2}{8}.
\end{align*}
Notice that for any other set $S \subseteq [K]$ of the platforms that have entered the market, the above bound on $\alpha$ guarantees the existence of a platform equilibrium strategy as $|S| \le K$. This completes the proof. $\blacksquare$
\subsection*{Proof of \cref{Pro:Equilibirum:K=1}}
The user utility  from sharing if the platform enters the market is
\begin{align*}
    \mathcal{I}_1 - \alpha \mathcal{I}(\sigma_1)= \frac{1}{2} - \alpha \frac{\gamma_1^2}{2+ \sigma_1^2}.
\end{align*}
The above utility is always positive if $\alpha \le 1/\gamma_1^2$. Therefore, the user always shares their data. This also implies that the platform chooses $\sigma_1=0$ to maximize the payment received from the buyer. The platform's utility then becomes 
\begin{align*}
    \mathcal{I}_1 - c_1 + \beta \mathcal{I}(0)= \frac{1}{2} - c_1 + \beta \frac{\gamma_1^2}{2}.
\end{align*}
Therefore, the platform enters the market if and only if the above quantity is positive.

If $\alpha > 1/\gamma_1^2$, then the user's utility is not always positive. In this case, if the platform enters the market, it chooses $\sigma_1$ so that the user utility is zero, i.e., 
\begin{align*}
    \mathcal{I}_i - \alpha \mathcal{I}(\sigma_1)=\frac{1}{2} - \alpha \frac{\gamma_1^2}{2+ \sigma_1^2}=0.
\end{align*}
The above equality gives us $\sigma_1= \sqrt{2 \left( \alpha \gamma_1^2 -1 \right)}$. The platform's utility then becomes 
\begin{align*}
    \mathcal{I}_1 - c_1 + \beta \mathcal{I}(\sigma_1)= \frac{1}{2} - c_1 + \beta \frac{1}{2 \alpha}.
\end{align*}
Therefore, the platform enters the market if and only if the above quantity is positive. $\blacksquare$
\subsection*{Proof of \cref{Cor:welfare}}
When all platforms enter the market, as we have established in Lemma \ref{Lem:BoundaryofSigma1}, the equilibrium noise variance is such that the user is indifferent between sharing with all platforms and not sharing at all. This means that 
\begin{align}\label{eq:pf:Cor:zeroutility:1}
    \sum_{i=1}^K \mathcal{I}_i - \alpha \mathcal{I}(\sigma_1, \dots, \sigma_K)=0.
\end{align}
Moreover, the payments cancel out in the utilitarian welfare and it becomes 
\begin{align*}
    2 \sum_{i=1}^K \mathcal{I}_i + \left(\beta - \alpha \right) \mathcal{I}(\sigma_1, \dots, \sigma_K) - \sum_{i=1}^K c_i= K +\left(\beta - \alpha \right) \frac{K}{2 \alpha} - \sum_{i=1}^K c_i,
\end{align*}
where the above equality follows from $\mathcal{I}_i=\frac{1}{2}$ for all $i \in [K]$ and \eqref{eq:pf:Cor:zeroutility:1}. 

The argument is similar to the above one when only low-cost platforms enter the market. $\blacksquare$

\subsection*{Proof of \cref{pro:uniform:mandate:versus:ban}}
If we only have low-cost platforms, then all platforms enter the market as their utility of  platform $i \in [K]$ is given by 
\begin{align*}
    \mathcal{I}_i - c_i + p_i = \frac{1}{2} - c_i + p_i \ge 0,
\end{align*}
which is always positive. Notice that if we ban data sharing, then the user shares with all platforms (because there is no privacy loss), and its utility becomes 
\begin{align*}
    \sum_{i=1}^K \mathcal{I}_i= \frac{K}{2}.
\end{align*}
Clearly, there is no uniform minimum privacy mandate that can make the user better off because banning data sharing maximizes the user's gain from services and minimizes the privacy loss. 

Now, let us prove the second part of this proposition. By banning data sharing, only low-cost platforms enter the market, and the user shares data with all of them and obtains the following utility:
\begin{align*}
    \frac{1}{2} (\text{ number of low-cost platforms}).
\end{align*}
Now consider a uniform minimum privacy mandate $\bar{\sigma}$ such that 
\begin{align*}
    \frac{1}{2} K - \alpha \mathcal{I}(\underbrace{\bar{\sigma}, \dots, \bar{\sigma}}_{K \text{ many}} ) > \frac{1}{2} (\text{ number of low-cost platforms}).
\end{align*}
For any $\alpha$, as $\bar{\sigma}$ increases, the leaked information goes to zero, and therefore, so long as there exists at least one high-cost platform, a $\bar{\sigma}$ for which the above inequality holds exists. We next argue that all platforms enter the market for a large enough $\beta$. To see this, we need to have 
\begin{align*}
    \mathcal{I}_i- c_i + \beta \Bigl(\mathcal{I}(\underbrace{\bar{\sigma}, \dots, \bar{\sigma}}_{K \text{ many}} ) - \mathcal{I}(\underbrace{\bar{\sigma}, \dots, \bar{\sigma}}_{K-1 \text{ many}} ) \Bigr) \ge 0
\end{align*}
for all $i \in [K]$. Therefore, letting $\beta \ge \tilde{\beta}$ where 
\begin{align*}
    \tilde{\beta} = \max\Bigl\{\bar{\beta},  \frac{c_i- \frac{1}{2}}{\mathcal{I}(\underbrace{\bar{\sigma}, \dots, \bar{\sigma}}_{K \text{ many}} ) - \mathcal{I}(\underbrace{\bar{\sigma}, \dots, \bar{\sigma}}_{K-1 \text{ many}} )}\Bigr\}
\end{align*}
and $\bar{\beta}$ is specified in \cref{thm:equilibirum:K>2} all platforms enter the market. $\blacksquare$
\subsection*{Proof \cref{pro:optimal:mandate}}
Similar to the proof of \cref{pro:uniform:mandate:versus:ban} and \cref{thm:equilibirum:K>2}, we choose $\bar{\alpha}$ and $\hat{\beta}$ large enough so that an equilibrium exists and all platforms enter the market. Now, suppose $\bar{\sigma}$ is the optimal uniform minimum privacy mandate so that all platforms enter the market, the user utility is
\begin{align*}
    \frac{1}{2}K -\alpha \mathcal{I}(\underbrace{\bar{\sigma}, \dots, \bar{\sigma}}_{K \text{ many}} ),
\end{align*}
and that the utility of platform $i$ is 
\begin{align*}
    \mathcal{I}_i-c_i + \beta \Bigl(\mathcal{I}(\underbrace{\bar{\sigma}, \dots, \bar{\sigma}}_{K \text{ many}} ) - \mathcal{I}(\underbrace{\bar{\sigma}, \dots, \bar{\sigma}}_{K-1 \text{ many}} ) \Bigr) \ge 0.
\end{align*}
Consider a regulation that bans data sharing for low-cost platforms and uses the same minimum privacy mandate for high-value platforms. We first prove that all platforms still enter the market. First, notice that the low-cost platforms enter the market as they obtain a positive utility. Letting $h$ be the number of high-cost platforms, the utility of a high-cost platform $i$ becomes 
\begin{align*}
    \mathcal{I}_i-c_i + \beta \Bigl(\mathcal{I}(\underbrace{\bar{\sigma}, \dots, \bar{\sigma}}_{h \text{ many}} ) - \mathcal{I}(\underbrace{\bar{\sigma}, \dots, \bar{\sigma}}_{h-1 \text{ many}} ) \Bigr) \ge  \mathcal{I}_i-c_i + \beta \Bigl(\mathcal{I}(\underbrace{\bar{\sigma}, \dots, \bar{\sigma}}_{K \text{ many}} ) - \mathcal{I}(\underbrace{\bar{\sigma}, \dots, \bar{\sigma}}_{K-1 \text{ many}} ) \Bigr) \ge 0,
\end{align*}
where the first inequality follows from the submodularity of the leaked information in actions, established in \cref{Lem:submodular}. The utility of the user also increases because we have 
\begin{align*}
    \frac{1}{2}K -\alpha \mathcal{I}(\underbrace{\bar{\sigma}, \dots, \bar{\sigma}}_{h \text{ many}} ) \ge \frac{1}{2}K -\alpha \mathcal{I}(\underbrace{\bar{\sigma}, \dots, \bar{\sigma}}_{K \text{ many}} ),
\end{align*}
where the inequality follows from the monotonicity of the leaked information, established in \cref{Lem:monotone}. $\blacksquare$

{
\section{Extensions and Additional Results}\label{App:Extensions}

\subsection{Extension to Non-linear Utilities}\label{section:extension_nonlinear}
Here, we establish that our main results and insights continue to hold in a setting in which the utility of the agents is not necessarily linear in the leaked information. In particular, in line with \eqref{eqn:utility_user}, we let the user utility be
\begin{equation} \label{eqn:utility_user:extend}
\mathcal{U}_{\text{user}}(\bsigma, \be, \ba, \bp, \bb):= 
\sum_{i=1}^K a_i  e_i \mathcal{I}_i
- \alpha ~ H_u\left(\mathcal{I}(\bsigma, \be, \ba, \bb) \right),
\end{equation}
for some function $H_u(\cdot)$ which is strictly increasing and concave:
\begin{align}\label{ass:1}
    \inf_{x \in [0,1]} H'_u(x) >0, ~~~ H'_u(\cdot) \text{ is decreasing}, ~~~ \text{ and } H_u(0)=0.
\end{align}
Similarly, in line of \eqref{eqn:utility_buyer}, the 
data buyer's utility is given by
\begin{equation}\label{eqn:utility_buyer:extend}
\mathcal{U}_{\text{buyer}}(\bsigma, \be, \ba, \bp, \bb):= 
\beta ~ H_b\left(\mathcal{I}(\bsigma, \be, \ba, \bb) \right) - \sum_{i=1}^K b_i p_i,
\end{equation}
for some function $H_b(\cdot)$ which is strictly increasing and concave:
\begin{align}\label{ass:2}
    \inf_{x \in [0,1]} H'_b(x) >0, ~~~ H'_b(\cdot) \text{ is non-increasing}, ~~~ \text{ and } H_b(0)=0.
\end{align}
The utility of the platform is the same as the one given in \eqref{eqn:utility_platform_i}.
In all the utilities, we keep the term $\mathcal{I}_i$ linearly because this is a constant term, and applying a function to it gives us another constant without substantially changing the game. 

Our main result in Theorem \ref{thm:equilibirum:K>2} continue to hold:

\noindent \textbf{Theorem 1':}
Suppose Assumptions \ref{assumption:substitute} and \ref{assumption:all_regions} hold. Additionally, suppose \eqref{ass:1} and \eqref{ass:2} hold. There exists $\bar{\alpha}$, $\bar{\beta}$, and $\underline{\beta}$ such that for $\alpha \ge \bar{\alpha}$, we have the following:
\begin{itemize}
    \item If $\beta \ge \bar{\beta}$, an equilibrium of the game exists. Moreover, in any equilibrium, all platforms enter the market, the user utility is zero, and the buyer purchases the data of all platforms and has a strictly positive utility.
    \item If $\beta \le \underline{\beta}$, an equilibrium of the game exists. Moreover, in any equilibrium, only low-cost platforms enter the market, the user utility is zero, and the buyer purchases the data of low-cost platforms and has a strictly positive utility only if there is more than one low-cost platform.
\end{itemize}

In what follows, we establish the proof of this theorem and, along the way, state how each of the results from the main text continues to hold in this setting. 

\noindent\textbf{Lemma 3':} For a given vector of noise levels $\bsigma$, the revealed information is submodular in the set of platforms that share the user data with the data buyer, i.e., for any $\bsigma$, $S \subseteq T \subseteq [K]$, and $i \not\in T$, we have 
\begin{align*}
    H_b\left(\mathcal{I}((\sigma_i, \bsigma_S)) \right)- H_b \left(\mathcal{I}(\bsigma_S) \right) \ge H_b\left(\mathcal{I}((\sigma_i, \bsigma_T)) \right)-H_b\left(\mathcal{I}(\bsigma_T) \right). 
\end{align*}
\noindent\emph{Proof:} This follows from a similar argument to that of Lemma \ref{Lem:submodular} by noting that the above inequality can rewritten as
\begin{align*}
    \int_{\mathcal{I}(\bsigma_S) }^{\mathcal{I}((\sigma_i, \bsigma_S))} H'_b(s) ds \ge \int_{\mathcal{I}(\bsigma_T)}^{\mathcal{I}((\sigma_i, \bsigma_T))} H'_b(s) ds.
\end{align*}
Now, by the monotonicity of the revealed information, the left-hand side integral starts at a lower point, and by Lemma \ref{Lem:submodular}, the interval over which we integrate is longer. These two facts, together with the assumption that $H'_b(s)$ is non-negative and decreasing, complete the proof. $\blacksquare$

\noindent \textbf{Proposition 1':}
For a given $(\bsigma, \be, \ba)$, there exists a unique price equilibrium given by 
\begin{align*}
    p_i^*= \beta \left(H_b\left( \mathcal{I}(\bsigma, \be, \ba, \mathbf{1}) \right) - H_b\left(\mathcal{I}(\bsigma, \be, \ba, (b_i=0,\mathbf{1}_{-i}))\right) \right) \text{ for all } i \in [K]
\end{align*}
and the corresponding unique buyer equilibrium is $b_i=1$ for all $i \in [K]$, where $\mathbf{1}_{-i}$ is the vector of all ones for all $j \in [K] \setminus \{i\}$. 

\emph{Proof:} It follows from the same argument as that of Proposition \ref{Pro:Equilibirum:Buyer:Price}. $\blacksquare$

\noindent\textbf{Lemma 4':} For any $\bsigma$, $S \subseteq [K]$, $i \not\in S$, and $j \in S$
\begin{align*}
    H_b\left(\mathcal{I}((\sigma_i, \bsigma_S)) \right)-H_b \left(\mathcal{I}(\bsigma_S) \right)
\end{align*}
is decreasing in $\sigma_i$ and increasing in $\sigma_j$.

\noindent\emph{Proof:} The above difference can be written as
\begin{align*}
    \int_{\mathcal{I}(\bsigma_S)}^{\mathcal{I}((\sigma_i, \bsigma_S))} H'_b(s) ds,
\end{align*}
which is decreasing in $\sigma_i$ and increasing in $\sigma_j$ because of Lemma \ref{Lem:submodular:sigma} and that $H'_b(\cdot)$ is non-negative and decreasing. $\blacksquare$

\emph{Proof of Theorem 1':} Similarly, we can see that Lemma \ref{Lem:BoundaryofSigma1} continues to hold in this setting. Following the proof of Theorem \ref{thm:equilibirum:K>2}, the proof proceeds to hold by letting 
\begin{align*}
    \bar{\beta}:= \sup_{i \in [K], (e_i=1, \be_{-i}), \bsigma \in \Sigma_{\mathbf{1}, \be} \cap \Sigma_{\mathbf{0}, \be}} \frac{c_i - \frac{1}{2}}{H_b\left(\mathcal{I}(\bsigma, \be, (a_i=1, \ba_{-i}), \mathbf{1})\right) -H_b\left( \mathcal{I}(\bsigma, \be, (a_i=0, \ba_{-i}), \mathbf{1})\right) }
\end{align*}
and 
\begin{align*}
    \underline{\beta}:= \inf_{i \in [K], c_i>\frac{1}{2}, (e_i=1, \be_{-i}), \bsigma \in \Sigma_{\mathbf{1}, \be}} \frac{c_i - \frac{1}{2}}{H_b\left(\mathcal{I}(\bsigma, \be, (a_i=1, \ba_{-i}), \mathbf{1})\right) - H_b\left( \mathcal{I}(\bsigma, \be, (a_i=0, \ba_{-i}), \mathbf{1})\right) }.
\end{align*}
Therefore, a similar argument to that of Theorem \ref{thm:equilibirum:K>2} shows that if an equilibrium exists, then depending on whether $\beta \ge \bar{\beta}$ or $\beta \le \underline{\beta}$ we have the cases listed in the statement of Theorem 1'. Thus, it remains to prove that for large enough $\alpha$, an equilibrium exists. Again, we use the same argument as that presented in the proof of Theorem \ref{thm:equilibirum:K>2} and Lemma \ref{Lem:closed-form:g}, and, in particular, show that for large enough $\alpha$, $(\sigma_1^2, \cdots, \sigma_K^2)$ such that
\begin{align*}
    \frac{\gamma_i}{2+ \sigma_i^2} = t \text{ for all } i\in [K] 
\end{align*}
where $t$ is such that 
\begin{align}\label{eq:pf:Thm1p:t}
    \frac{K}{2}=\alpha H_u\left( \mathcal{I}(\sigma_1, \dots, \sigma_K)\right) =\alpha H_u \left( \frac{K t}{1+ (K-1)t} \right)
\end{align}
is an equilibrium. 

First, note that, as $\alpha$ grows, the $t$ for which the above identity holds goes to zero. In fact, using the Taylor expansion (along with $H_u(0)=0$), it is straightforward to verify that $t$ is given by
\begin{align} \label{eqn:approx_t}
t= t_{o} + o\left(\frac{1}{\alpha} \right) 
\quad 
\text{with }
t_o := \frac{1}{1+2 \alpha H'_u(0)-K}.
\end{align}
Next, to show why the proposed noise levels constitute an equilibrium. Similar to the proof of Theorem \ref{thm:equilibirum:K>2}, let us consider the deviation of platform 1 so that the corresponding user equilibrium (after deviation) is $(a_1=1, \ba_{-i})$ with $\ba_{-i}$ having $i-1$ ones and $K-i$ zeros. 
Therefore, we have
\begin{align}\label{eq:pf:Thm1p:tp}
    \frac{K}{2} - \alpha H_u \left(\mathcal{I}\left(\tilde{\sigma}, \sigma_2, \cdots,  \sigma_K \right) \right) \leq \frac{i}{2} - \alpha \tilde{H}_u\left(\mathcal{I}\left( \tilde{\sigma}_1, \sigma_2, \cdots, \sigma_i \right) \right).
\end{align}
Let us define $t'=\frac{\gamma_1^2}{2+ \tilde{\sigma}_1^2}$. Note that the noise after deviation, i.e., $\tilde{\sigma}_1$, is larger than $\sigma_1$ (as otherwise platform 1 will get no data at user equilibrium as the user equilibrium was already at the boundary of sharing and not sharing with all platforms). Moreover, the revealed information only decreases after deviation, and it is already $\mathcal{O}(1/\alpha)$ before deviation. Therefore, by the Taylor expansion, we have
\begin{align*}
\frac{K}{2} - \alpha H'_u(0) \mathcal{I}(\tilde{\sigma}_1, \dots ,\sigma_K) \leq  \frac{i}{2} - \alpha H'_u(0) \mathcal{I}(\tilde{\sigma}_1, \dots ,\sigma_i) + o \left( \frac{1}{\alpha} \right),
\end{align*}
where $o \left( \frac{1}{\alpha} \right)$ represents the higher-order terms. We can rewrite this as
\begin{align*}
\frac{K-i}{2} \leq  \alpha H'_u(0) 
\Big (\mathcal{I}(\tilde{\sigma}_1, \dots ,\sigma_K) - 
\mathcal{I}(\tilde{\sigma}_1, \dots ,\sigma_i)  \Big) + o \left( \frac{1}{\alpha} \right).
\end{align*}
Next, by invoking Lemma \ref{Lem:closed-form:g}, we can substitute the exact formulation for the above revealed information as a function of $t$ and $t'$, as we did in the proof of \Cref{thm:equilibirum:K>2}. This way, we obtain
\begin{align} \label{eqn:bound_t_prime_1}
\frac{K-i}{2} \leq  H'_u(0)
\Big ( \frac{(K-i)(1-t) \alpha t}{\left(1+t'+(i-2)t \right) \left(1+t'+(K-2)t \right)}
\Big) + o \left( \frac{1}{\alpha} \right).
\end{align}
Now, if we ignore the $o \left( \frac{1}{\alpha} \right)$ term in \eqref{eqn:bound_t_prime_1}, similar to the proof of \Cref{thm:equilibirum:K>2}, we have:
\begin{align} \label{eqn:t_prime_UB_2}
t'\leq t'_o:= \frac{2-4 \alpha H'_u(0) + K-i + \sqrt{16 \alpha^2 H'_u(0)^2+(K-i)^2-8 \alpha H'_u(0) K}}{2+4 \alpha H'_u(0) - 2 K}.
\end{align}
Next, let us see how much $t'$ can go up if we consider the term $o \left( \frac{1}{\alpha} \right)$. Note that, since $\alpha t$ is constant, it can be verified that the derivative of the term
\begin{equation*}
\frac{(K-i)(1-t) \alpha t}{\left(1+t'+(i-2)t \right) \left(1+t'+(K-2)t \right)}    
\end{equation*}
from \eqref{eqn:bound_t_prime_1} has a constant lower bound, and therefore, we
have
\begin{align} \label{eqn:t_prime_UB_3}
t'\leq t'_o + o \left( \frac{1}{\alpha} \right).
\end{align}
Now, we are ready to establish why such a deviation is indeed not profitable for platform 1. To do so, we need to show the following inequality holds:
\begin{align*}
 H_b \left(\mathcal{I}(\sigma_1, \cdots, \sigma_K)\right)  - H_b \left(\mathcal{I}(\sigma_2, \cdots, \sigma_K)\right) \ge H_b \left(\mathcal{I}(\tilde{\sigma}_1, \cdots, \sigma_i)\right) - H_b \left(\mathcal{I}(\sigma_2, \cdots, \sigma_i)\right).
\end{align*}
Invoking Lemma \ref{Lem:closed-form:g}, we need to prove
\begin{align}\label{eq:pf:Thm1p:dev}
 & H_b \left( \frac{Kt}{1+(K-1)t}  \right)  - H_b\left( \frac{(K-1)t}{1+(K-2)t} \right) \\
 & \ge H_b \left(1- \frac{1-t}{1+t'+(i-2)t}  \right)- H_b \left(\frac{(i-1)t}{1+(i-2)t}  \right). 
\end{align}
Note that the right hand side is increasing in $t'$ as $H_b(\cdot)$ is an increasing function, and therefore, it suffices to show this bound for the upper bound \eqref{eqn:t_prime_UB_3}. 
Also, recall from \eqref{eq:pf:Thm1p:t} that $\frac{Kt}{1+(K-1)t}$ is $\mathcal{O}(1/\alpha)$.
Therefore, by Taylor expansion, it suffices to prove
\begin{align}\label{eq:pf:Thm1p:dev_2}
 &  \frac{Kt}{1+(K-1)t}  - \frac{(K-1)t}{1+(K-2)t} 
 \ge 1- \frac{1-t}{1+t'+(i-2)t}  - \frac{(i-1)t}{1+(i-2)t} + E, 
\end{align}
where $E$ is the combined error of the Taylor approximation, which we know is bounded by $o \left( \frac{1}{\alpha} \right)$. Next, we substitute $t$ from \eqref{eqn:approx_t} and the upper bound on $t’$ from \eqref{eqn:t_prime_UB_3}. Using the fact that all terms have bounded derivatives, it suffices to prove that
\begin{align}\label{eq:pf:Thm1p:dev_3}
 &  \frac{Kt_o}{1+(K-1)t_o}  - \frac{(K-1)t_o}{1+(K-2)t_o} 
 \ge 1- \frac{1-t_o}{1+t'_o+(i-2)t_o}  - \frac{(i-1)t_o}{1+(i-2)t_o} + E', 
\end{align}
where $E'$ represent all error terms and is order of $o \left( \frac{1}{\alpha} \right)$. Note that the left-hand side minus the right-hand side is what we characterized in the proof of \Cref{thm:equilibirum:K>2}, modulo replacing $\alpha$ by $\alpha H'_u(0)$, and is given by \eqref{eqn:diff_two_sides_1}, which we have shown is lower bounded by
\begin{equation} \label{eqn:diff_two_sides_2}
\frac{3K-1}{4 \alpha H'_u(0)}- \frac{K-1}{2 \alpha H'_u(0)}  - \frac{K}{2 \alpha H'_u(0) \left(1+ \sqrt{1- \frac{K}{2 \alpha H'_u(0)}} \right)}, 
\end{equation}
which is equal to
\begin{equation}\label{eqn:diff_two_sides_3}
\frac{1}{4\alpha H'_u(0)} \left( 
K+1 - \frac{2K}{1+ \sqrt{1- \frac{K}{2 \alpha H'_u(0)}}}
\right).    
\end{equation}
For \eqref{eq:pf:Thm1p:dev_3} to hold, we need this term to be bigger than $E'$, and so it suffices to show it is lower bounded by $c/\alpha$ for some constant $c$, as $E'$ is order of $o \left( \frac{1}{\alpha} \right)$. To do so, notice that, for 
\begin{equation*}
\alpha \geq \frac{(K+1/2)^2}{4 H'_u(0)},    
\end{equation*}
we have
\begin{equation}
K+1 - \frac{2K}{1+ \sqrt{1- \frac{K}{2 \alpha H'_u(0)}}} \geq \frac{1}{2},    
\end{equation}
which means \eqref{eqn:diff_two_sides_3} is lower bounded by $1/(8\alpha H'_u(0))$. This completes the proof. 
$\blacksquare$

\subsection{Extension to intermediary values of $\beta$ for $K>2$} \label{section:extension_intermidiary}
\Cref{thm:equilibirum:K>2} characterizes the equilibrium for sufficiently small and large values of $\beta$. In this section, we describe how the equilibrium behaves for intermediate values of $\beta$. Throughout this section, and without loss of generality, we assume that the costs of the different platforms satisfy the following order:
\begin{equation}
c_1 \leq \cdots \leq c_{\ell} \leq \frac{1}{2} < c_{\ell+1} < \cdots < c_K,    
\end{equation}
where $\ell$ denotes the number of low-cost platforms. To simplify the presentation of the results, we assume that all high-cost platforms have distinct costs. However, it is straightforward to extend the results to cases in which two or more platforms have identical costs.

We next present two main results, both concerning the order in which platforms enter the market as $\beta$ increases. We state the results for the case of linear utilities, but it is straightforward to verify that the techniques from the previous section can be extended to the nonlinear case.

\begin{theorem} \label{theorem:intermidiary_exists}
Suppose Assumptions \ref{assumption:substitute} and \ref{assumption:all_regions} hold. For any $i \in \{\ell+1, \ldots, K\}$, define 
\begin{equation*}
\beta_i := \frac{2\alpha(2\alpha-1)}{(2\alpha-i)} \cdot \left (c_i - \frac{1}{2} \right ),    
\end{equation*}
with the convention that $\beta_\ell = 0$ and $\beta_{K+1} = \infty$. Then, there exists $\bar{\alpha}$, such that, for any $i \in \{\ell, \ldots, K\}$, any $\beta \in [\beta_i, \beta_{i+1})$, and any $\alpha \geq \bar{\alpha}$, there exists an equilibrium in which only platforms $1$ to $ i$ enter the market. The user’s utility is zero at equilibrium, and the buyer purchases all the data from the platforms that enter the market, obtaining strictly positive utility for $i \geq 2$. 
\end{theorem}
\emph{Proof:} 
Fix $i \in \{\ell, \ldots, K\}$. We consider the equilibrium constructed in the proof of \Cref{thm:equilibirum:K>2}, in which the first $i$ platforms enter the market with
\begin{equation}
\frac{\gamma_j}{2+\sigma_j^2} = t_i:=\frac{1}{1+2\alpha-i} \quad \forall j \in \{1,\cdots, i\}.   
\end{equation}
It remains to verify the market entry condition in \eqref{eqn:market_entering_condition}. In particular, by substituting the revealed information identity at equilibrium from \eqref{eqn:revealed_info_border}, together with the revealed information when one platform ceases sharing from \Cref{Lem:closed-form:g}, we can rewrite the condition for platform $j$ to enter the market as
\begin{equation}
\frac{1}{2} - c_j + \beta \left ( \frac{i}{2\alpha} - 1 + \frac{1-t_i}{1+(i-2)t_i} \right ) \geq 0. 
\end{equation} 
This further simplifies to 
\begin{equation}
\frac{1}{2} - c_j + \beta \frac{2\alpha-i}{2\alpha(2\alpha-1)} \geq 0.   
\end{equation}
Now, if $\beta \in [\beta_i, \beta_{i+1})$, this condition holds only for all $j \in \{1, \ldots, i\}$, which concludes the proof. $\blacksquare$

This result establishes that there exists a sequence of equilibria in which the low-cost platforms always enter the market, while the high-cost platforms enter sequentially, in order of their costs, as $\beta$ increases.

We complement this result with the following one, which shows that there exists a minimum value of $\beta$ representing the buyer’s willingness to pay in order to induce high-cost platforms to enter the market, and that this minimum increases with the platforms' costs.
\begin{proposition} \label{proposition:intermidiary_LB}
Suppose Assumptions \ref{assumption:substitute} and \ref{assumption:all_regions} hold. At any equilibrium in which a high-cost platform with cost $c$ enters the market, the following condition must hold:
\begin{equation*}
\beta \geq 2\alpha (c-\frac{1}{2}).    
\end{equation*}
\end{proposition}
\emph{Proof:} 
Suppose we have an equilibrium with platforms' strategy vector $(\bsigma, \be)$, in which $i$ platforms enter the market, including the high-cost platform $j$. Then, by \eqref{eqn:market_entering_condition}, we should have
\begin{equation} \label{eqn:beta_LB_1}
\frac{1}{2}-c_j + \beta \left(\mathcal{I}(\bsigma, \be, (a_j=1, \ba_{-j}), \mathbf{1}) - \mathcal{I}(\bsigma, \be, (a_j=0, \ba_{-j}), \mathbf{1}) \right) \ge 0.    
\end{equation}
Now, we know by \eqref{eqn:revealed_info_border} that 
\begin{equation} \label{eqn:beta_LB_2}
\mathcal{I}(\bsigma, \be, (a_j=1, \ba_{-j}), \mathbf{1}) = \frac{i}{2 \alpha}.    
\end{equation}
Moreover, by \Cref{Lem:BoundaryofSigma1}, we have
\begin{equation}
\frac{i}{2} - \alpha \mathcal{I}(\bsigma, \be, (a_j=1, \ba_{-j}), \mathbf{1}) \geq 
\frac{i-1}{2} - \alpha \mathcal{I}(\bsigma, \be, (a_j=0, \ba_{-j}), \mathbf{1}).
\end{equation}
By \eqref{eqn:beta_LB_2}, the left-hand side is equal to zero, which implies
\begin{equation} \label{eqn:beta_LB_3}
 \mathcal{I}(\bsigma, \be, (a_j=0, \ba_{-j}), \mathbf{1}) \geq \frac{i-1}{2\alpha}.   
\end{equation}
Plugging \eqref{eqn:beta_LB_2} and \eqref{eqn:beta_LB_3} into \eqref{eqn:beta_LB_1} completes the proof. $\blacksquare$

}
\end{document}